\documentstyle[psfig]{mn}

\begin{document}

\journal{To Appear in MNRAS}

\title[$\Omega_0$ and substructure in galaxy clusters]{
$\mathbf{\Omega_0}$ and substructure in galaxy clusters}

\author[David A. Buote]{David A. Buote\thanks{E-mail:
buote@ast.cam.ac.uk} \\ Institute of Astronomy, Madingley Road,
Cambridge CB3 0HA}

\maketitle

\begin{abstract}

We examine the theoretical relationship between $\Omega_0$ and
substructure in galaxy clusters which are formed by the collapse of
high density peaks in a gaussian random field.  The radial mass
distributions of the clusters are computed from the spherical
accretion model using the adiabatic approximation following Ryden \&
Gunn.  For a cluster of mass, $M(r,t)$, we compute the quantity
$\Delta M /\, \overline{M}$ at a cosmic time t and within a radius
$r$, where $\Delta M$ is the accreted mass and $\overline{M}$ is the
average mass of the cluster during the previous relaxation time, which
is computed individually for each cluster.  For a real cluster in
three dimensions we argue that $\Delta M /\, \overline{M}$ should be
strongly correlated with the low order multipole ratios,
$\Phi^{int}_l/\Phi^{int}_0$, of the potential due to matter interior
to $r$. Because our analysis is restricted to considering only the low
order moments in the gravitational potential, the uncertainty
associated with the survival time of substructure is substantially
reduced in relation to previous theoretical studies of the ``frequency
of substructure'' in clusters.

We study the dependence of $\Delta M /\, \overline{M}$ on radius,
mass, $\Omega_0$, $\lambda_0=1-\Omega_0$, redshift, and relaxation
timescale in universes with Cold Dark Matter (CDM) and power-law power
spectra. The strongest dependence on $\Omega_0$ $(\lambda_0=0)$ occurs
at $z=0$ where $\Delta M /\, \overline{M}\propto\Omega_0^{1/2}$ for
relaxation times $\sim 1-2$ crossing times and only very weakly
depends on mass and radius.  The fractional accreted mass in CDM
models with $\Omega_0~+~\lambda_0~=~1$ depends very weakly on $\Omega_0$
and has a magnitude similar to the $\Omega_0=1$ value.  $\Delta M /\,
\overline{M}$ evolves more rapidly with redshift in low-density
universes and decreases significantly with radius for $\Omega_0=1$
models for $z\ga 0.5$.  We discuss how to optimize constraints on
$\Omega_0$ and $\lambda_0$ using cluster morphologies.

It is shown that the expected correlation between $\Delta M /\,
\overline{M}$ and $\Phi^{int}_l/\Phi^{int}_0$ extends to the
two-dimensional multipole ratios, $\Psi^{int}_m/\Psi^{int}_0$, which
are well defined observables of the cluster density distribution.  We
describe how N-body simulations can quantify this correlation and thus
allow $\Delta M /\, \overline{M}$ to be measured directly from
observations of cluster morphologies.

\end{abstract}

 
\section{Introduction}
\label{intro}

Clusters of galaxies have proven to be useful laboratories for
cosmological studies and have in particular yielded interesting
measurements of $\Omega_0$, the present value of the average mass
(energy) density of the universe expressed in terms of the critical
density required for closure. The methods to measure $\Omega_0$ with
clusters each have their own advantages and disadvantages (see Dekel,
Burstein, \& White 1997).  Perhaps the most familiar method is to
assume the mass-to-light ratio in clusters is representative of the
universe as a whole \cite{cye}. This method is popular because of its
conceptual simplicity and relative ease to implement, but the basic
assumption of this method may be unjustified.  Another promising
method uses the abundances of clusters (e.g. White, Efstathiou, \&
Frenk 1993) to derive a relationship between $\Omega_0$ and
$\sigma_8$, the rms fluctuations in mass within spheres of radius
$8h^{-1}$ Mpc. This method has the advantage that the theoretical
comparison can be usefully achieved by relatively simple and
computationally inexpensive semi-analytical methods.  One disadvantage
of this method is that cluster masses are usually determined from
X-ray temperatures, the relationship of which is not entirely certain.
Another disadvantage is the uncertainty in $\sigma_8$ due the
uncertainty in the bias factor for clusters \cite{mjw}.

The detailed structure of clusters provides a complementary constraint
on $\Omega_0$. Although the radial density profiles of clusters appear
to be very insensitive to $\Omega_0$ when scaled in terms of their
virial radii (Navarro, Frenk, \& White 1997), the presence of
substructure in the mass distribution appears to be a quite sensitive
diagnostic of $\Omega_0$ as suggested in the pioneering study by
Richstone, Loeb, \& Turner \shortcite{rlt}.  Advantages of this method
include the direct influence of $\Omega_0$ on cluster morphologies
(i.e. growth of structure regulated by $\Omega_0$), and cluster
morphologies are straightforward to quantify and compute
observationally (e.g. Buote \& Tsai 1995).  The principal
disadvantage is that a proper theoretical model requires
computationally expensive high-resolution N-body simulations of a {\it
large} number of clusters sufficient to adequately sample the variety
of spatial morphologies of a cluster population (Jing et al. 1995;
Buote \& Xu 1997; Thomas et al. 1997).

To date the only study that has compared the requisite large N-body
simulations to observations is Buote \& Xu (1997; hereafter BX).
Using dissipationless simulations of scale-free and Cold Dark Matter
(CDM) models (e.g. Ostriker 1993) and X-ray data of clusters from
ROSAT, BX determined that the spatial morphologies of galaxy clusters
(1) decouple the influence of $\Omega_0$ and the power spectrum when
quantified in terms of statistics based on ratios of multipoles of the
projected gravitational potential (Buote \& Tsai 1995), and (2) favor
a low matter density $(\Omega_0\la 0.3)$.  However, a rigorous
determination of $\Omega_0$ from X-ray cluster morphologies requires
dissipational simulations of a statistically large number of clusters,
at present a daunting computational expenditure even for only one set
of model parameters\footnote{Dissipationless simulations should be
adequate for mass maps of clusters derived from gravitational lensing
when such maps for observed clusters become widely available.}.  To
facilitate this process, less computationally intensive semi-analytic
methods could be used to rapidly identify the optimum parameters
(e.g. redshift, mass range, radius) of a cluster sample which maximize
the ability to determine $\Omega_0$ and possibly a cosmological
constant via cluster morphologies.

A re-examination of semi-analytical\footnote{For our discussion we
broadly define semi-analytical to be techniques other than
three-dimensional N-body simulations.} models of the relationship
between cluster morphologies and $\Omega_0$ is important in its own
right.  As it is to be expected given the nature of the problem,
semi-analytic approaches to cluster morphologies have been highly
idealized.  The previous studies do not actually construct model
cluster mass distributions having substructure, but either adopt a
spherical collapse model for computing a distribution of cluster
collapse times \cite{rlt} or use simplified models of the merging
histories of clusters to compute the distributions of times since a
substantial merger event (Kauffmann \& White 1993; Lacey \& Cole
1993).  Even if we accept such simplifications as necessary evils to
be traded for intuitive guidance and increased computational speed,
these models are of limited usefulness because they only compute the
ambiguous ``frequency of substructure'' in clusters which is very
sensitive to the uncertain survival time of substructure (Kauffmann \&
White 1993; Lacey \& Cole 1993).

In this paper we present an intuitive model for substructure, based on
the spherical accretion model, that is closely related to quantitative
observables of cluster morphology.  We describe how N-body simulations
can quantify the relationship between our model and the observables
and thus allow a direct measurement of $\Omega_0$. However, for our
present study, we use this model to (1) provide physical insight into
the evolution of cluster morphology with redshift as a function of
$\Omega_0$ and $\lambda_0=1-\Omega_0$ and to (2) assist the designing
of future observing programs for the detailed comparison of these
observational samples to clusters produced in N-body simulations.  We
then outline a procedure to optimize determination of $\Omega_0$ and
$\lambda_0$ using the observables of cluster morphology that are
closely related to our model.

The paper is organized as follows. In \S \ref{mot} we motivate our
model and show its relationship to quantitative observables of cluster
morphology. The details of the spherical accretion model and the
cosmological framework are discussed in \S \ref{computation} and \S
\ref{cospar} respectively. In \S \ref{results} we analyze the model
for different values of $\Omega_0$, $\lambda_0$, and $z$ for CDM and
power-law power spectra. Finally, we discuss these results and
identify the best ways to constrain $\Omega_0$ and $\lambda_0$
observationally in \S \ref{disc}.

\section{Motivation}
\label{mot}

\subsection{Theory}
\label{theory}

In a standard Friedmann-Robertson-Walker universe with $\Omega_0<1$
and $\lambda_0=0$, the linear growth of density fluctuations becomes
strongly suppressed when the curvature term in the Friedmann equation
exceeds the matter term.  The redshift delineating this transition
from an Einstein - de Sitter phase to one of free expansion is then
$1+z_{trans}=\Omega_0^{-1}-1$; i.e. when the matter density
$\Omega(z_{trans})=0.5$. (e.g. \S 11B in Peebles 1980). Hence, if
$\Omega_0\ll 1$, then objects formed a long time in the past relative
to universes where $\Omega_0\approx 1$, and thus clusters in
low-density universes should be, on average, more relaxed than
clusters in universes with $\Omega_0\approx 1$.  To apply this idea to
real clusters it is necessary to specify what it means for clusters to
be ``more relaxed''.

The most general stable self-gravitating non-rotating equilibrium
configuration is the triaxial ellipsoid. Features like substructure
that break this symmetry provide a measure of the departure of a
cluster from a virialized state.  The relaxation rate determines how
rapidly substructure and other non-ellipsoidal features are erased in
order to bring the system to equilibrium.  Hence, the amount of
substructure in a cluster at a particular epoch, or rather the degree
of non-ellipsoidal symmetry, is approximately determined by the amount
of mass accreted over the timescale associated with the relaxation
rate.

This relaxation timescale depends on the mass distributions of the
cluster and the subclumps (e.g. White \& Rees 1978; Binney \& Tremaine
1987), and is typically of order the crossing time of the system for
substructures and non-ellipsoidal distortions comprising a substantial
fraction $(\ga 20\%)$ of the cluster mass\footnote{Mergers of
structures of these sizes with clusters are typical for clusters
formed in N-body simulations (e.g. Tormen, Bouchet, \& White 1997).}.
Since our interest lies in how much a cluster departs from a
virialized state, we must consider how much a cluster evolves
morphologically over the relaxation timescale, not just the amount of
substructure present at a particular time.

If gravity drives the dynamical evolution of clusters, then we would
expect the morphological evolution to be similar for clusters of
different masses (approximately self-similar); i.e. the degree of
morphological evolution should be proportional to the amount of
accreted mass and inversely proportional to the total cluster mass.
We write this fractional amount of mass accreted over a crossing time
as $\Delta M /\, \overline{M}$, where $\Delta M$ is the accreted mass
and $\overline{M}$ is the average mass of the cluster within a radius
$r$.  We may write an equivalent expression in terms of the multipole
moments of the gravitational potential due to matter interior to a
radius $r$,
\begin{equation}
{\Delta M \over \overline{M}} = {\Delta\Phi^{int}_0\over
\overline{\Phi}^{int}_0}, \label{eqn.def} 
\end{equation}
where $\Phi^{int}_0= -GM(<r)/r$ is the monopole term.  Of course,
substructure and other non-ellipsoidal distortions have important
contributions to other moments of the potential, and we expect that
the increase in the monopole will be strongly correlated with
increases in the next few multipoles. (Indeed, this is the basis of
our hypothesis that substructure is related to the amount of accreted
mass over the relaxation timescale.)  Hence, let us consider the
increase in $\Phi^{int}$, expanded in terms of all of its multipoles
(and suppressing the azimuthal terms),
\begin{equation}
{\Delta\Phi^{int}\over\overline{\Phi}^{int}} = { {\sum_l
\alpha_l\Phi^{int}_l}\over \overline{\Phi}^{int}}, \label{eqn.mult}
\end{equation}
where $\Phi^{int}_l$ is the $l$th term in the multipole expansion at
time $t$ of interest, and $\alpha_l$ characterizes the increase in
$\Phi^{int}_l$ over the previous crossing time; i.e. $\alpha_l=0$ if
the particular multipole does not increase, and $\alpha_l=1$ if the
multipole term is zero at time $t-t_{cross}$ and non-zero at $t$.  For
our discussion we will consider a spherical region positioned at the
center of mass\footnote{The center of mass can be defined within the
sphere of radius $r$ through iteration.} and thus we do not require
the azimuthal terms in the multipole expansion -- although our general
argument does not depend sensitively on the shape of the aperture used
to compute the moments.  Since the $l>0$ terms are not spherically
symmetric, we average $\Phi^{int}_l$ over the spherical surface of
radius $r$ and denote it by $\langle\Phi^{int}_l\rangle$. Actually we
average $(\Phi^{int}_l)^2$ because $\langle\Phi^{int}_l\rangle = 0$
for $l>0$.

Because we cannot observe a cluster for the duration of a crossing
time, let us approximate equation (\ref{eqn.mult}) at the time $t$ of
interest.  First, in the denominator we set
$\overline{\Phi}^{int}\approx\Phi^{int}$ which is accurate to within a
factor of 2, and set $\Phi^{int}\approx\Phi^{int}_0$, since the
monopole term dominates the higher order terms for all reasonable
cases.  For a relaxed, smooth cluster the $l>0$ ratios
$\langle(\Phi^{int}_l)^2\rangle/\langle(\Phi^{int}_0)^2\rangle$ are
substantially smaller than for a cluster with subclumps that are a
considerable $(\ga 20\%)$ fraction of the total mass (Buote \& Tsai
1995)\footnote{Buote \& Tsai actually discuss circularly averaged
moments of the projected potential which replace the cluster mass
density with toy models of the X-ray surface brightness of
clusters. Nevertheless, the qualitative features apply when the mass
is used as shown \S 4 of Tsai \& Buote \shortcite{tb}.}.  As a result,
we may approximate equation (\ref{eqn.mult}) by setting $\alpha_l=1$
for the $l>0$ terms
\footnote{This approximation implies that
$\langle(\Delta\Phi^{int})^2\rangle/\langle(\overline{\Phi}^{int})^2\rangle\ne
0$ for virialized clusters because the even-$l$ moments, though small
in relation to clusters with considerable substructure, are non-zero
in general for ellipsoidal masses.}.  Making these substitutions we
arrive at,
\begin{equation}
{\langle(\Delta\Phi^{int})^2\rangle\over\langle(\overline{\Phi}^{int})^2\rangle}
\approx \left({\Delta M \over \overline{M}}\right)^2 + 
 \sum_{l>0} {\langle({\Phi^{int}_l})^2\rangle \over
\langle(\Phi^{int}_0)^2\rangle}, 
\label{eqn.key} 
\end{equation}
which states that the fractional increase in the rms spherically
averaged potential over a crossing time is approximately the
fractional increase in the mass and the increases in the ratios of the
rms spherically averaged higher order multipoles to the monopole added
in quadrature. Violent relaxation \cite{lb} is the key process driving
the elimination of large potential fluctuations. It operates on a
timescale of $\sim 1-2$ crossing times and proceeds independently of
the masses of the constituents in accordance with our picture of the
morphological evolution of clusters stated above.

Hence, equation (\ref{eqn.key}) is a definition of the dynamical state
of a cluster and could have been the starting point of our
discussion\footnote{For alternative indicators of the dynamical state
of a cluster see Zaroubi, Naim, \& Hoffman \shortcite{z} and
Natarajan, Hjorth, \& van Kampen \shortcite{priya}.}.  By the nature
of the multipole expansion, we see that only the low order moments are
important for characterizing the dynamical state, with the monopole
term being most important\footnote{Our formulation is a twist on the
argument of Richstone et al. (1992) who computed the fraction of
present-day clusters which ``collapsed'' over the previous relaxation
time. This fraction is then interpreted as the ``frequency of
substructure'' in the present cluster population.  In our case, for an
individual cluster the fractional accreted mass over the relaxation
time approximately determines the importance of substructure in the
cluster and thus its dynamical state.}.  As a result, the possible
long-term survival of the dense, compact cores of subclusters, which
contribute mostly to the higher order moments, does not affect the
shorter relaxation timescale of the low order moments most relevant to
the dynamical state.  (The implications of this property are discussed
in \S\S \ref{results} and \ref{disc}.)

\subsection{Observational Consequences}
\label{obs}

The fractional change in the monopole, $\Delta M /\, \overline{M}$,
cannot be directly observed over a relaxation time.  However, our
premise that the amount of accreted mass over the previous relaxation
timescale determines the amount of substructure (or non-ellipsoidal
distortions) requires that the monopole change be strongly correlated
with the change in the low order terms, which are approximately the
ratios,
$\langle(\Phi^{int}_l)^2\rangle/\langle(\Phi^{int}_0)^2\rangle$,
defined at the epoch under consideration.  What is the nature of this
correlation and, more importantly, how may $\Delta M /\, \overline{M}$
be inferred from cluster observations?

For observations of clusters it is convenient to work with the
projected potential, $\Psi^{int}$, and its circularly averaged (over
radius $R$) multipole ratios,
$\langle(\Psi^{int}_m)^2\rangle/\langle(\Psi^{int}_0)^2\rangle$.  For
the $m>0$ terms it is unclear how to relate a measured
$\langle(\Psi^{int}_m)^2\rangle/\langle(\Psi^{int}_0)^2\rangle$ to the
multipole ratios in three-dimensions. However, the fractional change
of the monopoles in 2-D and 3-D are comparable if the fractional
accreted masses within the 2-D and 3-D regions are comparable.  That
is, let us again consider our prototype case where the 3-D region is a
sphere of radius $r$ defined at its center of mass. The corresponding
2-D region is the circle of radius $R=r$ resulting from the projection
along its symmetry axis of the cylinder which encloses the sphere at
its center.  So long as the extra accreted mass in the portion of the
cylinder outside the sphere does not contribute much to the total
fractional accreted mass of the cylinder, then the 2-D and 3-D
fractional monopole changes will be similar.

Hence, ${\Delta\Phi^{int}_0 /\, \overline{\Phi}^{int}_0}\cong
{\Delta\Psi^{int}_0 /\, \overline{\Psi}^{int}_0}$ for projected radii
containing a sizeable fraction of the total cluster mass. (An
appropriate minimum radius can be determined from N-body simulations.)
If we now consider the 2-D analogue of equation (\ref{eqn.key}), then
the correspondence of the 2-D and 3-D monopole changes coupled with
the arguments made in the preceding section indicate that $(\Delta M
/\, \overline{M})^2$ (3-D) should be strongly correlated with
$\langle(\Psi^{int}_m)^2\rangle/\langle(\Psi^{int}_0)^2\rangle$, which
are themselves easily computable observables of the cluster density
distribution (Buote \& Tsai 1995).  For observations of an individual
cluster this correspondence assumes that projection does not smear out
the structure; e.g. axisymmetric clusters are viewed edge-on. Later
in this section we discuss how to deal with projection
effects\footnote{We show in \S \ref{cdmz0} that the
$\langle(\Psi^{int}_m)^2\rangle/\langle(\Psi^{int}_0)^2\rangle$
obtained from the N-body simulations of BX agree reasonably well with
our calculations for $\Delta M /\,\overline{M}$.}.

Three dimensional N-body simulations can explicitly quantify the
relationship between $\Delta M /\,\overline{M}$ and
$\langle(\Psi^{int}_m)^2\rangle/\langle(\Psi^{int}_0)^2\rangle$.
There is reason to believe that this relationship will be very tight
for small $m$ and mostly independent of $\Omega_0$.  For clusters
observed with {\it ROSAT} (Buote \& Tsai 1996) and those formed in
both dissipationless (BX; Thomas et al. 1997) and dissipational N-body
simulations \cite{tb}, the ratios,
$\langle(\Psi^{int}_m)^2\rangle/\langle(\Psi^{int}_0)^2\rangle$, are
strongly correlated for the first few $m$ -- in fact the logarithms of
these ratios are essentially proportional to each other
\footnote{Again, although most of the references listed have focused
on moments of the X-ray surface brightness, Tsai \& Buote (1996)
showed that taking moments of the mass gives very consistent
results.}. We henceforth adopt the simplified notation $P_m/P_0\equiv
\langle(\Psi^{int}_m)^2\rangle/\langle(\Psi^{int}_0)^2\rangle$ used by
these other studies.

Tsai \& Buote \shortcite{tb} studied these correlations in detail for
both the X-ray surface brightness maps and the projected masses of six
clusters formed in the N-body / hydrodynamical simulation of Navarro,
Frenk, \& White \shortcite{nfw95}.  These correlations can be
understood as tracks followed by clusters as they evolve via mergers
and subsequent relaxation from infancy to quasi-virialized states (see
Figure 6 of Tsai \& Buote and \S 5.1 of Buote \& Tsai 1996). For the
larger values of $P_m/P_0$ (corresponding to the least relaxed
clusters) the $m=2,3,4$ ratios are tightly correlated and are very
nearly proportional to each other.  

The scatter in these correlations is largest for the smallest values
of $P_m/P_0$ as a result of projection effects.  The smallest values
of $P_m/P_0$ correspond to either relaxed, single-component clusters
or to clusters with substructure smeared out by the act of projecting
along the line of sight; e.g. a bimodal cluster being viewed along the
merger axis.  Thus, to minimize the projection effects the
correlations between the low order $P_m/P_0$ (particularly for
$\log_{10}P_2/P_0$ versus $\log_{10}P_4/P_0$) can be defined
accurately for the larger values of $P_m/P_0$ where the scatter is low
(see, e.g., Figure 3 of Buote \& Tsai 1996, Figure 4 of BX, and Figure
9 of Thomas et al. 1997.) Moreover, the approximation $\alpha_l\equiv
1$ in equation (\ref{eqn.key}) is most rigorously satisfied for these
largest multipole ratios.

The shapes of these correlations do not show significant differences
as a function of $\Omega_0$ or $\lambda_0=1-\Omega_0$ (BX), although
Thomas et al. \shortcite{thomas} note that the $\log_{10}P_2/P_0$
versus $\log_{10}P_4/P_0$ correlation is tighter for the $\lambda_0=0$
models. Since these correlations are expected to apply as well to
${\Delta M/\, \overline{M}}$ (3-D), we may write,
\begin{equation}
\log_{10}\left({\Delta M \over \overline{M}}\right)^2 = 
c_m\log_{10}\frac{P_m}{P_0} + d_m,
\label{eqn.dmcorr}
\end{equation}
where $c_m$ and $d_m$ are the linear coefficients which should be
largely independent of $\Omega_0$ and, for best accuracy, should be
determined from fits to the largest values of $\log_{10}P_m/P_0$. By
quantifying the $c_m$ and $d_m$ with N-body simulations we may infer
$\Delta M /\, \overline{M}$ from observations of $P_m/P_0$ for a large
sample of clusters and thus measure the value of $\Omega_0$\footnote{A
minimum of about 20 clusters appears to be required to distinguish
between CDM models with $\Omega_0=1$ and $\Omega_0=0.3$
\cite{thomas}.}.  (Another advantage of writing the expected
correlations in terms of logarithms is that BX found that the means of
the $\log_{10} P_m/P_0$ ($m=2,3,4$) distributions of N-body clusters
are influenced primarily by $\Omega_0$ whereas the variances are
affected mostly by the power spectrum of density fluctuations;
i.e. the influence of the power spectrum may be largely circumvented
by examining the means of the logarithmic distributions.)

Measurements of $P_m/P_0$ for clusters may be obtained from either
gravitational lensing or X-ray data (for a discussion see \S 2 of
Buote \& Tsai 1995)\footnote{Another route to observations is via
cooling flows (e.g. Fabian 1994), where the cooling-flow rate should
be {\it anti-correlated} with $\Delta M /\, \overline{M}$
\cite{bt2}}. Since gravitational lensing measures the projected mass
directly, the coefficients $c_m$ and $d_m$ can be quantified by
dissipationless simulations which is an advantage over X-ray
data. However, the projection of the mass in simulated clusters
appears to lead to smaller measured variations in $P_m/P_0$ than does
the X-ray surface brightness (see \S 4 of Tsai \& Buote 1996), and
thus the larger dynamic range afforded by X-ray maps probably
indicates a greater sensitivity to $\Omega_0$. This larger dynamic
range occurs because for unrelaxed clusters the X-ray emitting gas
approximately traces the mass but its emission is more responsive to
density fluctuations since it goes as $\rho_{gas}^2$. In the other
limit, when the gas is in hydrostatic equilibrium the emission traces
the potential for arbitrary temperature gradient (\S 3.1 of Buote \&
Canizares 1994; \S 5.1 of Buote \& Canizares 1996) and is thus
smoother and rounder than the mass (see \S 2.3.1 of BX for a detailed
discussion).

\subsection{Present Application}
\label{present}

Presenting a constraint on $\Omega_0$ by obtaining the coefficients
$c_m$ and $d_m$ (equation \ref{eqn.dmcorr}) from N-body simulations
for comparison to observations is beyond the scope of the present
paper.  Instead we prefer to focus on the {\em differences} between
CDM models with different $\Omega_0$ since the differences in $\Delta
M /\, \overline{M}$ can be related to differences in $P_m/P_0$ without
knowledge of the as yet unknown coefficients $d_m$ and by taking
$c_m\approx 1$ as suggested by the $\log_{10}P_m/P_0$
correlations. For this task we can compute the quantity $\Delta M /\,
\overline{M}$ using a spherical model for a cluster which has the
significant advantages of being simpler conceptually and considerably
less intensive computationally than three-dimensional N-body
simulations.  In this way we can gain physical insight into the
morphological evolution of clusters in different cosmologies and
rapidly explore the interesting parameter space to aid analysis of
future three dimensional simulations and observations.

Of course, a spherical model is only an idealization, but the radial
density profiles seen in three dimensional N-body simulations agree
well with simple spherical accretion models (e.g. Navarro, Frenk, \&
White 1995; Anninos \& Norman 1996; also see \S \ref{masspro} of this
paper). By using a spherical model we in essence compute a mean value
of $\Delta M /\, \overline{M}$, whereas in a more realistic model the
non-spherical nature of the merging process will induce fluctuations
in ${\Delta M /\, \overline{M}}$ depending on how the mass is
distributed in clumps during a merger.  However, for studying the
dependence of $\Delta M /\, \overline{M}$ on $\Omega_0$, the mean
value should be sufficient since BX showed that it is primarily
$\Omega_0$ which determines the means of the distributions of
$P_m/P_0$ for clusters formed in three-dimensional N-body simulations.
We therefore adopt a spherical model for our present investigation.

\section{Computation of $\mathbf\Delta M /\, \overline{M}$}
\label{computation}

\subsection{Peaks formalism}
\label{peaks}

It is our intention to work within the standard framework wherein
small density inhomogeneities in the early universe grow by
gravitational instability into the galaxy clusters that we see today.
We restrict our discussion to the case where these initial density
fluctuations are described by a gaussian random field, and, in
particular, that the sites where clusters form are determined by high
peaks in this perturbation field (Bardeen et al. 1986, hereafter
BBKS).  With these assumptions the average initial density profile
around such high peaks is specified by the power spectrum of density
fluctuations, $P(k)$. (For similar presentations of the following see
Ryden \& Gunn 1987, Ryden 1988, Hoffman 1988, and especially Lilje \&
Lahav 1991.)

The spherical accretion model (see \S \ref{sphmod}) only depends on
the density through the initial cumulative density contrast,
$\overline{\delta}_i(x)$, defined at comoving position $x$ at initial
time $t_i$.  The initial cumulative density contrast around a high
peak in a gaussian random field is (BBKS),
\begin{equation}
\overline{\delta}_i = \left(\nu - {\gamma\theta\over 1-\gamma^2}\right)
{\overline{\xi}\over\sigma_0} - {\theta\over 1-\gamma^2}
{\overline{\nabla^2\xi} \over \sigma_2},
\end{equation}
where $\gamma\equiv\sigma^2_1/(\sigma_2\sigma_0)$ and the spectral
moments are defined by,
\begin{equation}
\sigma^2_{\mu} = {1\over 2\pi^2}\int P(k)k^{2+2{\mu}}dk,
\end{equation}
with the special case $\sigma^2_0 = \xi(0)$; the cumulative
correlation function and its cumulative Laplacian are,
\begin{eqnarray}
\overline{\xi} & = & {1\over 2\pi^2}\int P(k)k^2w(kx)dk\\
\overline{\nabla^2\xi} & = & {-1\over 2\pi^2}\int P(k)k^4w(kx)dk,
\end{eqnarray}
(note: $\overline{\nabla^2\xi}={3\over x}{d\xi\over dx}$) where the
top-hat filter is given by,
\begin{equation}
w(x) = {3\over x^3}\left(\sin x-x\cos x\right).
\end{equation}
The dimensionless parameter $\nu$ specifies the peak height,
$\delta_i(0) = \nu\sigma_0$, and $\theta(\gamma,\gamma\nu)$ is given
by equation (6.14) of BBKS with the asymptotic behavior
$\theta\rightarrow 0$ as $\nu\rightarrow\infty$. Thus, for high peaks
$\overline{\delta} \approx \nu\overline{\xi}/\sigma_0$.  The peak only
dominates the collapse in its vicinity which we take to be defined as
the region where $\overline{\delta}_i\le \sqrt{
(\Delta\overline{\delta}_i)^2 }$, where the variance in
$\overline{\delta}_i$ is (BBKS),
\begin{equation}
(\Delta\overline{\delta}_i)^2 = \sigma^2_M - {1\over 1-\gamma^2}
\left[{\overline{\xi}^2\over\sigma^2_0} +
{\overline{\nabla^2{\xi}}\over\sigma_2}
\left(2\gamma{\overline{\xi}\over\sigma_0}+
{\overline{\nabla^2\xi}\over\sigma_2}\right)\right],
\end{equation}
where,
\begin{equation}
\sigma^2_M(x) = {1\over 2\pi^2}\int P(k)w^2(kx)k^2dk,
\end{equation}
is the mass variance in spheres of comoving radius $x$. 

We do not consider secondary perturbations arising from non-sphericity
and random velocities (BBKS) discussed by Ryden \& Gunn \shortcite{rg}
and Ryden \shortcite{ryden} in the context of galaxy-sized halos. Such
perturbations cause fluctuations in the collapse time of matter around
the peak, though we do not expect angular momentum to play a dominant
role in the formation of clusters since they are not observed to
rotate appreciably. At any rate, uncertainties of this variety that
affect the collapse time are related to our ignorance of the precise
relaxation timescale over which low-order potential fluctuations are
erased, the uncertainties of which we expect to dwarf those of the
secondary perturbations. (We discuss this issue further in the next
section.) 

For our calculations we define the initial epoch to be at
recombination, $z_i\equiv 1300$, though our results differ negligibly
for any $z_i\ga 50$.  In practice we compute $\overline{\delta}_i$ by
first evaluating $\overline{\delta}_0$, the cumulative density
contrast extrapolated to the present using linear theory, by
normalizing the linear power spectrum at $z=0$ (see \S
\ref{cospar}). Then $\overline{\delta}_i$ is computed from
$\overline{\delta}_0$ assuming linear growth, for which we use the
convenient approximation of Carroll, Press, \& Turner \shortcite{cpt}
for a universe with matter density parameter $\Omega$ and density
parameter, $\lambda$, due to a cosmological constant.  

\subsection{Spherical accretion model}
\label{sphmod}

We evolve the initial cluster density distribution determined by the
peaks formalism to a final state using the spherical accretion
model\footnote{For a discussion of some of the caveats associated with
this type of approach see Lilje \& Lahav (1991) and Bernardeau (1994).}.
It will now prove convenient to work in proper coordinates, $r=ax$,
where $a(t)$ denotes the expansion factor at time $t$. For arbitrary
$\Omega_0$ and a monotonically decreasing density profile, the
spherical collapse model (e.g. Peebles 1980; Padmanabhan 1993; Sahni
\& Coles 1995) exactly describes the expansion of spherical shells out
to their maximum radius (``turn around'').  (In the Appendix we give
the equations for the zero-curvature universe with a cosmological
constant $\lambda_0=1-\Omega_0$.)  After the shell reaches maximum
expansion, it collapses and crosses the orbits of other shells at
which point the spherical model no longer accurately describes the
motion of the shell.  If one restricts the solutions to those that are
self-similar, then exact solutions may be found (Fillmore \& Goldreich
1984; Bertschinger 1985).  For the general case, however, methods must
be used which specifically treat the interactions between the
shells. The standard method to do this is with N-body simulations
(e.g. Sigurdsson, Hernquist, \& Quinlan 1995; Thoul \& Weinberg
1995).

Another approach is to consider the growth of a density peak by
computing the orbit of an infalling shell in the potential generated
by the previously collapsed matter (Gunn 1977; Ryden \& Gunn 1987 --
also see Blumenthal et al. 1986).  If the infalling shell induces only
a small change in this potential, then the shell's orbit may be
computed using the adiabatic approximation. The accuracy of this
approximation is determined by the number of shells, $N$, used to
divide up the mass distribution. Typically, only for the first few
(innermost) shells does the potential change substantially. For large
$N$ ($\sim 100$) the errors associated with this approximation have a
negligible effect on shells at radii that are of interest to our
present study.

To construct a cluster using the adiabatic method (see Ryden \& Gunn
1987) one begins with a core mass distribution, taken to correspond to
the scale over which the power spectrum is smoothed (see \S
\ref{cospar}). The details of the shape of this core have no tangible
effect on the shells at radii of interest.  We consider now the first
mass shell orbiting in the potential generated by the core mass. The
energy per unit mass of the shell is an integral of the motion and is
equal to the potential at the maximum radius of the orbit. More
generally, we may write for a shell $n$ falling into the potential,
$\Phi_{n}$, generated by the mass of the previous $n-1$ shells plus
the core as,
\begin{eqnarray}
E_n & = & \Phi_{n}(r_m^n) \\
    & = & \frac{1}{2}  \left(\frac{dr}{dt}\right)^2 + \Phi_{n}(r),
\end{eqnarray}
where $r_m^n$ is the (proper) maximum radius of the $n$th shell. For
the first shell we have $\Phi_1 = \Phi_c$, the potential generated by
the core mass. The amount of mass a shell contributes within a radius
interval $dr$ is determined by the fractional amount of time that it
spends passing through $dr$ during its orbit.  The amount of time the
shell spends within a radius interval $dr$ is just $dt=dr/v_r$, and
thus the fractional amount of time, which is just the probability of
finding the shell within $dr$, is,
\begin{equation}
P_n(r)dr = { { dr/v^n_r} \over {\int_0^{r_m^n} {dr/v^n_r}} },
\end{equation}
and $v^n_r = |dr/dt|$ is the radial speed of shell $n$, which for the
first shell is,
\begin{equation}
v^1_r = \sqrt{2[E_1-\Phi_c(r)]}.
\end{equation}
Hence, the combined mass of the core and the first shell may be
written generally,
\begin{equation}
M_n(r) = M_{n-1}(r) + M_{sh}\int_0^r P_n(r)dr,
\end{equation}
where $M_{sh}$ is the total mass of the shell and $M_0=M_c$.

The next shell is added analogously to the first by computing its
energy, $E_2$, in the potential, $\Phi_2$, generated by mass, $M_1$,
and results in a new mass, $M_2$.  Now, however, the first shell sees
a different mass due to the overlapping orbit with the second
shell. If the potential change is small, then the energy of the first
shell decreases, but since the radial action,
\begin{equation}
j_r = \int_0^{r_m}v_rdr,
\end{equation}
is conserved, the effect is to reduce the maximum radius of the orbit
of the first shell \cite{g77}\footnote{For the first few shells the
radii can actually increase because the density profile changes
drastically in shape with the addition of each new shell; i.e. the
addition of the first few shells is not really adiabatic if the shell
mass is comparable to the core mass.}.  Hence, when the second shell
is added, $j_r$ is recomputed for the first shell in the potential
generated by the mass $M_2$. Then $r_m^1$ is adjusted so that the
action equals its initial value.  For an adiabatic change in the
potential, the new value of $r_m$ is approximately,
\begin{equation}
r_m^{new} = r_m^{old}{j_r^{old}\over j_r^{new}}, \label{eqn.actiter} 
\end{equation}
where $j_r^{new} = \int_0^{r_m^{old}}v_r^{new}dr$, with $v_r^{new}$
computed using the potential incorporating the new overlapping shell
-- for the case under consideration, the potential generated by the
mass $M_2$. Equation (\ref{eqn.actiter}) can be iterated by computing
a new $j_r^{new}$ using $r_m^{new}$ as the new upper limit in the
integral until desired precision in $r_m^{new}$ is achieved. Once the
radius of the first shell has been modified to reflect the addition of
the second shell, a new value of $M_2$ is computed.

(At this point one should in principle repeat this procedure for the
first and second shells until $r_m^1$ and $M_2$ change within desired
tolerances. However, corrections of this type are negligible when the
initial potential change is small, which applies as $n$ becomes
large. We neglect these higher order corrections.)

This adiabatic addition of shells continues until all of the collapsed
shells have been incorporated into the aggregate virialized
cluster. In general not all of the mass bound to the peak has
collapsed at the redshift of interest, say $z$.  We define the last
collapsed shell to be that which has just reached $r=0$ at redshift
$z$. We compute this directly by integrating $dr/v_r$ for each shell
rather than estimating the collapse time as $2t_m$. For the shells
that are bound to the peak but uncollapsed at $z$, we let them fall
into the cluster potential and assign them to their infall radii at
$z$.  Note that Ryden \& Gunn (1987; Ryden 1988) did not treat the
uncollapsed matter while Hoffman \shortcite{hof} allowed all of the
matter to collapse.

After building the cluster mass distribution at the epoch of interest,
we evaluate the crossing time which we take to be,
\begin{equation}
t_{cross} \equiv 2{R_{1/2}\over \sigma}, \label{eqn.tcross}
\end{equation}
where $R_{1/2}$ is the half-mass radius and $\sigma$ is the
one-dimensional (radial) velocity dispersion computed using the virial
theorem.  Next, we construct the cluster mass distribution at the
earlier time, $t-t_{cross}$. Finally, we evaluate the fractional
amount of mass accreted within a radius $r$ over the time interval
$(t-t_{cross},t)$,
\begin{equation}
{\Delta M \over \overline{M}} = {M(r,t) - M(r,t-t_{cross}) \over
\frac{1}{2}[M(r,t) + M(r,t-t_{cross})]},
\end{equation}
where $\Delta M$ is the mass difference and $\overline{M}$ is the
average mass.  As we show in \S \ref{results}, the value of ${\Delta M
/\, \overline{M}}$ is quite sensitive to the time interval
$(t-t_{cross},t)$ used to represent the relaxation timescale.  We need
only consider a relaxation timescale that is $\sim 1-2$ crossing times
since, as mentioned in \S \ref{mot}, we are interested only in
substructure contributing to fluctuations in the low order moments of
the gravitational potential.

However, even the relatively modest range of $\sim 1-2$ crossing times
causes sufficiently large changes in ${\Delta M /\, \overline{M}}$ to
make further refinements to our spherical accretion model unwarranted;
e.g. the secondary perturbations due to non-sphericity and angular
momentum discussed in \S \ref{sphmod}, initial peculiar velocities
\cite{bes}, or a small (and difficult to precisely define) drag force
due to substructure (Antonuccio-Delogu \& Colafrancesco 1994; Del
Popolo \& Gambera 1996).  Each of these effects either increase or
decrease the collapse times for the mass shells, of which the combined
effect is unclear, and such uncertainties should be largely absorbed
into the $\sim 1-2$ crossing time range assumed for the relaxation
timescale (as is the uncertainty in our specific definition of
$t_{cross}$ itself).

We now describe a quantity related to ${\Delta M /\, \overline{M}}$
which does not require following the detailed virialization process of
the cluster. The mass shell that has just collapsed at the epoch of
interest, $t$, reached maximum expansion approximately at a time,
$t/2$.  For a given peak height, we can invert the equation for the
time at maximum expansion to obtain the initial radius of this shell.
The mass enclosed by this shell is,
\begin{equation}
M(<r_i) =
\frac{4\pi}{3}\rho_b(t_i)r_i^3\left(1+\overline{\delta}_i\right),
\end{equation}
where $r_i$ is the initial (proper) radius. Because at maximum
expansion the shell has not yet crossed the orbits of other shells,
the enclosed mass $M(<r_m) = M(<r_i)$, where $r_m$ is the radius of
maximum expansion.  Let us denote this enclosed mass as $M_{ta}(t/2)
\equiv M(<r_m)$.  Analogously, we can estimate the mass that has
collapsed at the time $t-t_{cross}$ by computing
$M_{ta}([t-t_{cross}]/2)$. Hence, an estimate of the fractional
collapsed mass accreted over the previous crossing time is,
\begin{equation}
{\Delta M_{ta} \over \overline{M}_{ta}} = {M_{ta}(t/2) - M_{ta}([t-
t_{cross}]/2) \over
\frac{1}{2}\left[M_{ta}(t/2) + M_{ta}([t-t_{cross}]/2)\right]},
\label{eqn.mta} 
\end{equation}
which is computed directly from the turn-around times for the mass
shells and thus does not involve treating the detailed virialization
process. (The crossing time in this case is taken to be a constant
which is a good approximation for the clusters produced by the
adiabatic treatment of virialization -- see \S \ref{cdmz0}.)
Comparing ${\Delta M_{ta}/\, \overline{M}_{ta}}$ to ${\Delta M/\,
\overline{M}}$ allows us to assess the impact of incorporating the
virialization process into our calculations.

\section{Cosmological model parameters}
\label{cospar}

We consider power spectra appropriate for hierarchical clustering, in
particular CDM and power-law $P(k)$. We take the linear CDM power
spectrum according to BBKS,
\begin{eqnarray}
\lefteqn{P(k)_{CDM}\propto k\left({\log(1+2.34q) \over 2.34q}\right)^2
\times} \nonumber\\
& & \left[1 + 3.89q + (16.1q)^2 + (5.46q)^3 + (6.71q)^4\right]^{-1/2},
\end{eqnarray}
where we have assumed the scale-invariant form $(P(k)\propto k)$ for
the primordial spectrum. The spectrum is expressed in terms of the
parameter $q\equiv k/(h\Gamma)$, where the ``shape parameter'' is
defined to be \cite{sug},
\begin{equation}
\Gamma \equiv \Omega_0h\exp(-\Omega_B-\Omega_B\Omega_0^{-1}),
\label{eqn.gamma} 
\end{equation}
where $\Omega_B$ is the density parameter of baryonic mass.  We adopt
as our standard the value $\Omega_Bh^2=0.016$ \cite{copi} consistent
with Big Bang Nucleosynthesis. However, we also examine models with
the larger value inferred from the X-ray gas content in clusters,
$\Omega_Bh^{-3/2}/\Omega_0=0.05$ \cite{wf}.

We use the shape parameter to set the value of $h$ for the CDM models
($H_0=100h$ km s$^{-1}$ Mpc$^{-1}$). Viana \& Liddle \shortcite{vl}
determine 95\% confidence limits $\Gamma\approx 0.2-0.3$ using the
galaxy autocorrelation function data from Peacock \& Dodds
\shortcite{pd}.  We slightly expand this range to $\Gamma = 0.15-0.35$
and construct a linear relation with $\Omega_0$ such that $\Gamma =
0.15$ for $\Omega_0=0.2$ and $\Gamma = 0.35$ for $\Omega_0=1$. This
function $\Gamma(\Omega_0,h)$ -- at fixed $\Omega_B$ -- allows us to
determine a value of $h$ as a function of $\Omega_0$ consistent with
current data (though see Peacock 1997) with the slightly expanded
range chosen to keep the values of $h\approx 0.45-0.85$ within
reasonable observational limits (e.g. Fukugita, Hogan, \& Peebles
1993).

To eliminate the divergence of $P(k)_{CDM}$ at small wavelengths we
follow the standard procedure and mathematically smooth the power
spectrum with a gaussian filter, $P(k)_s \propto
P(k)_{CDM}\exp(-k^2l^2/2)$, where $l$ is the smoothing length in
comoving coordinates. The mass contained within the gaussian filter is
$M_s = (2\pi)^{3/2}\rho_b(l/\sqrt{2})^3$ at $z=0$.  Since we are
interested in mass fluctuations of the size of clusters of galaxies,
we choose $M_s=10^{12}h^{-1}M_{\sun}$ which corresponds to the mass of
a large galaxy at the present epoch, since structures with $M\la M_s$
individually have a negligible contribution to the dynamical evolution
of a cluster.  The smoothing length corresponding to this $M_s$ is $l
= 0.86\Omega_0^{-1/3}h^{-1}$ Mpc.  For evaluating ${\Delta M /\,
\overline{M}}$ at higher redshifts, we approximately account for the
smaller masses of galaxies by reducing the smoothing mass according
to, $M_s(z)=M_s(0)/(1+z)$, which is just the self-similar accretion
law $M\propto t^{2/3}$ \cite{edbert}.  (The results we obtain for
${\Delta M /\, \overline{M}}$ in the next section are not overly
sensitive to $l$.)

To normalize the smoothed CDM power spectrum we first set the value of
$\sigma_8 = \sigma_M(8h^{-1}\rm Mpc)$, the linearly extrapolated rms
mass fluctuation in spheres of radius $8h^{-1}$ Mpc, to agree with the
abundance of X-ray clusters at the present day $(z=0)$,
\begin{eqnarray}
\sigma_8 & = & 0.52 \Omega_0^{(-0.46 + 0.10\Omega_0)} \qquad\qquad
\lambda_0 = 0\nonumber \\ 
\sigma_8 & = & 0.52 \Omega_0^{(-0.52 + 0.13\Omega_0)} \qquad\qquad
\Omega_0 + \lambda_0 = 1, \label{eqn.sigma8}
\end{eqnarray}
where we have used the best-fit results of Eke, Cole, \& Frenk
\shortcite{ecf}. We obtain $\sigma_8$ for higher redshifts using the
same scaling relation for the density in  \S \ref{peaks}.  The
number of clusters as a function of mass measures the amplitude of
fluctuations in the mass on cluster scales (see, e.g., Frenk et
al. 1990), whereas we require the amplitude of fluctuations in
clusters (i.e. integrated over mass). In the peaks formalism, clusters
are biased tracers of the mass distribution which may be accounted for
by increasing the mass fluctuation amplitude by a bias factor (BBKS;
Bardeen et al. 1987),
\begin{equation}
b_{clus} \equiv { \langle \tilde{\nu}\rangle \over \sigma_0 } + 1,
\label{eqn.bias} 
\end{equation}
where $\langle \tilde{\nu}\rangle$ is an average peak height of the
model under consideration above some threshold $\nu_t$.  The cluster
power spectrum is then, $P(k)_{clus} = b_{clus}^2P(k)_{mass}$, where
$P(k)_{mass}$ is the smoothed CDM power spectrum normalized using the
above relation for $\sigma_8$.

We evaluate $b_{clus}$ in a manner similar to that described in Croft
\& Efstathiou \shortcite{ce} (also see Efstathiou et al. 1992). First,
we smooth the mass power spectrum with a gaussian of length $l =
5\sqrt{2}\Omega_0^{-1/3}h^{-1}$ Mpc and normalize to the above
$\sigma_8$. This smoothing length gives appropriate cluster masses
over a wide range of cluster abundances \cite{ce}. We focus our
attention on clusters with masses exceeding $3.5\times
10^{14}h^{-1}M_{\sun}$ corresponding to an abundance of $\sim 1\times
10^{-5}$ Mpc$^{-3}$ for typical CDM models (see Figure 1 of White,
Efstathiou, \& Frenk 1993). By requiring that the number density of
peaks in our model match this cluster abundance, we specify the peak
threshold, $\nu_t$, and thus $\langle \tilde{\nu}\rangle$. For CDM
models with $\Omega_0=0.2 - 1$ we obtain bias parameters
$b_{clus}\approx 1-3.5$.

We also consider models with power-law power spectra, $P(k)_{pl}
\propto k^n$, with $n$ ranging from 0 to -2.  Because $P(k)_{pl}$ also
diverges for small wavelengths we smooth it in the same manner as
$P(k)_{CDM}$. We normalize the smoothed power spectrum in the same
manner as done for the CDM spectrum and compute appropriate bias
factors. This is feasible because the $\sigma_8$-$\Omega_0$
relationship given above is very insensitive to the shape of the power
spectrum (e.g. White et al. 1993).

After choosing a power spectrum and specifying the remaining model
parameters, we construct clusters having masses spanning the range
$(0.35-3)\times 10^{15}h^{-1}M_{\sun}$. However, in the peaks
formalism the size of an over-density region is specified by the peak
height, $\nu$, rather than the mass.  When examining clusters at
$z>0$, we use $\nu$ determined at $z=0$; e.g. for a $1\times
10^{15}h^{-1}M_{\sun}$ cluster defined at $z=0$, we study its
progenitor at $z>0$ which has the same $\nu$ but smaller mass.  Note
that whenever we quote cluster masses in the next section we refer to
the actual mass computed from the full spherical model calculation
within $r=1.5h^{-1}$ Mpc.

\section{Results}
\label{results}

\subsection{Mass profiles at $\mathbf z=0$}
\label{masspro}

\begin{figure*}
\parbox{0.49\textwidth}{
\centerline{\psfig{figure=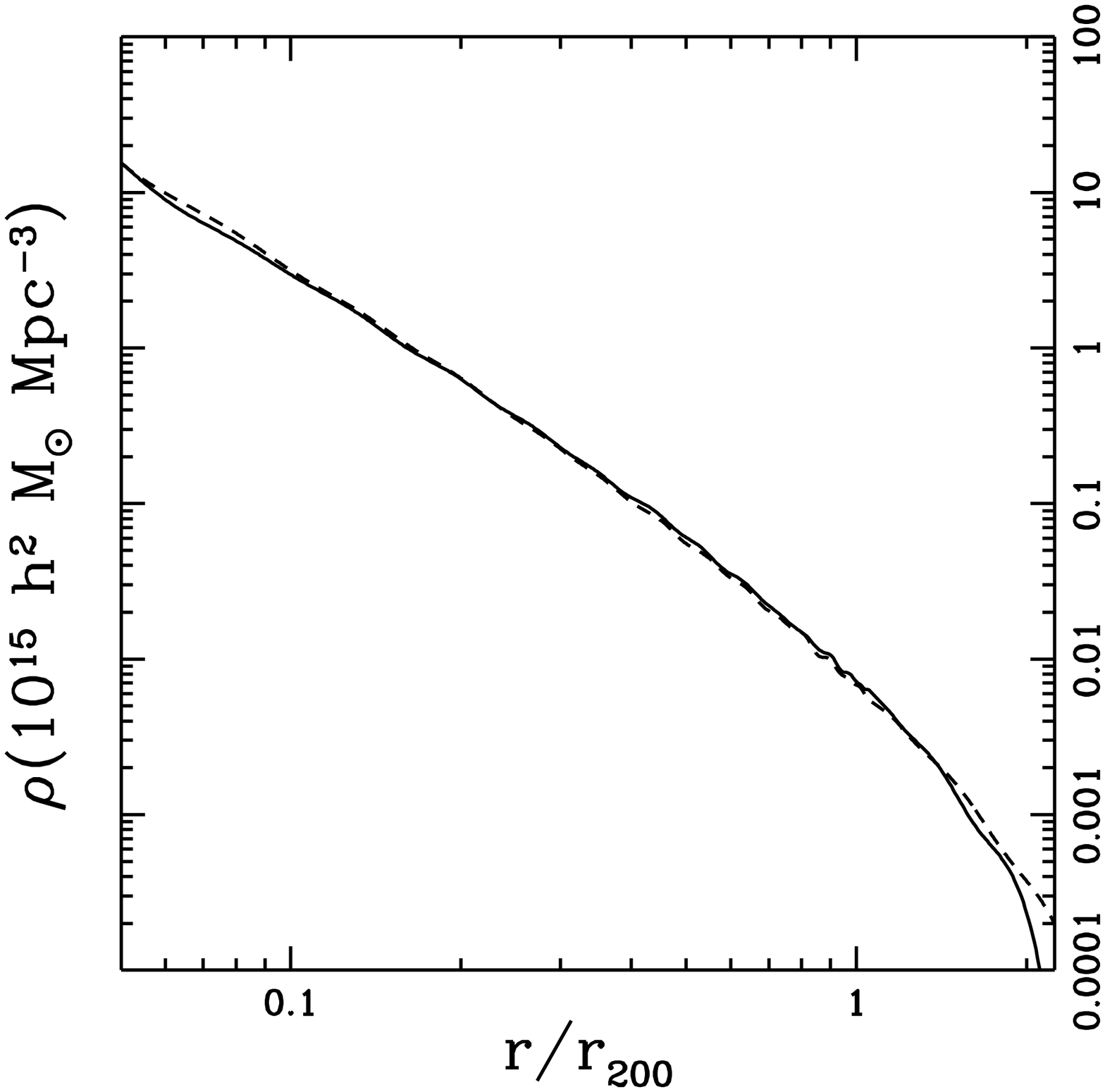,angle=0,height=0.3\textheight}}
}
\parbox{0.49\textwidth}{
\centerline{\psfig{figure=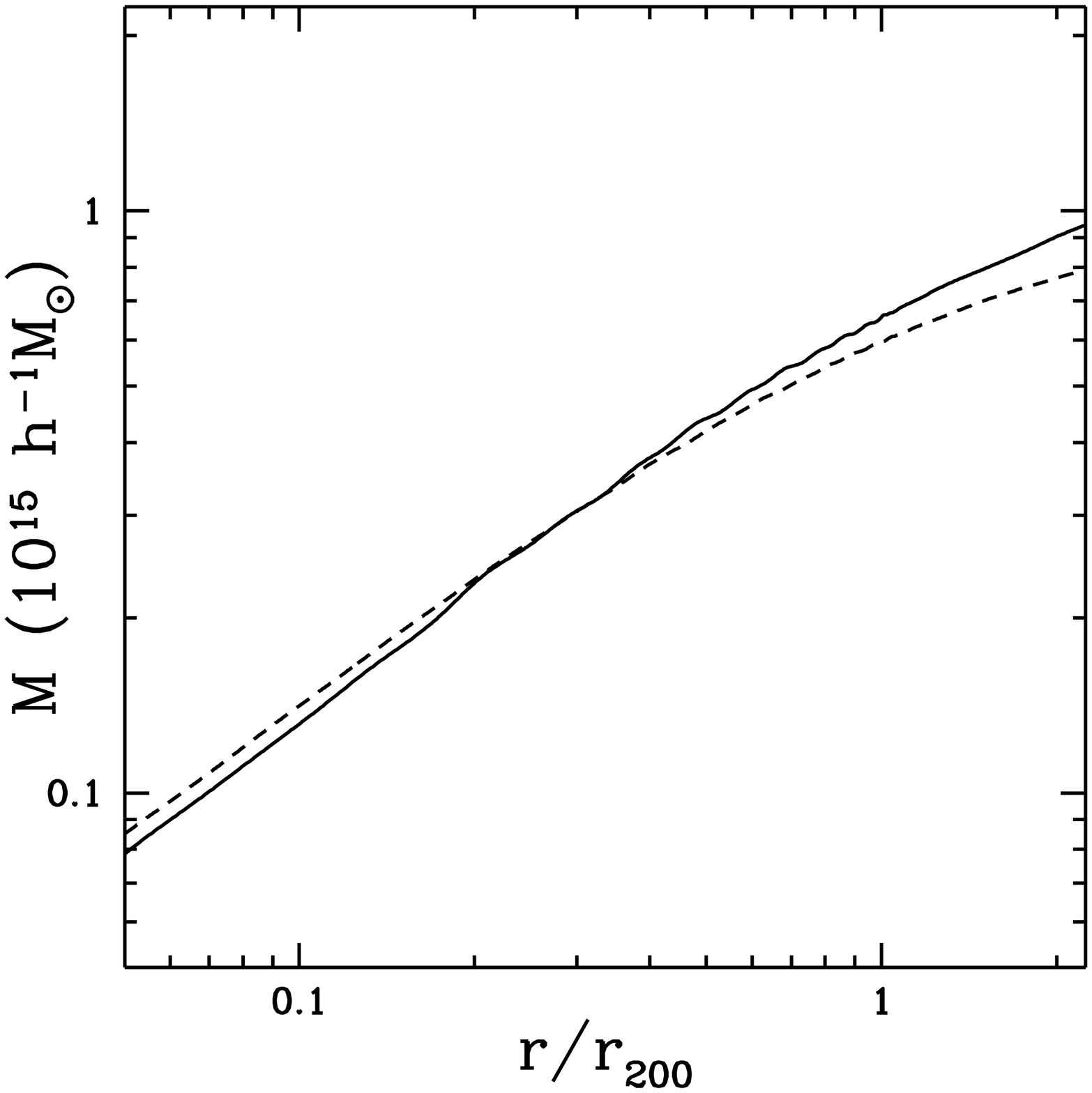,angle=0,height=0.3\textheight}}
}
\caption{\label{fig.denprof} Density (left) and mass (right) profiles
of $\approx 7\times 10^{14}h^{-1}M_{\sun}$ clusters in CDM models
with $\Omega_0=0.2$ (dashed) and $\Omega_0=1$ (solid). Only the
contribution from virialized matter is shown for the density profiles
while the mass profiles contain both the virialized and infalling
mass.}
\end{figure*}

Although not the focus of our present study, considerable literature
exists on the theoretical density profiles of clusters. Hence, to
place our model clusters in context, we briefly summarize the
properties of the cluster mass profiles obtained from the spherical
accretion model at $z=0$ for a CDM power spectrum. A more detailed
discussion of the density profiles will appear in a separate paper
\cite{bnew}.

In Figure \ref{fig.denprof} we display the density and mass profiles
of clusters with mass $\sim 7\times 10^{14}h^{-1}M_{\sun}$ formed in
models with $\Omega_0=0.2,1$.  Only the virialized matter is shown for
the density profiles since the infalling shells contribute (formally)
infinite density spikes at their locations. We express the radii in
terms of $r_{200}$, the radius where the mean cluster density is 200
times the background value: $r_{200}=1.63\times
10^{-5}(M/h^{-1}M_{\sun})^{1/3}$ Mpc (e.g. Navarro, Frenk, \& White
1997). For the clusters shown, $r_{200}\cong 1.5h^{-1}$ Mpc.

Without the infalling material $\rho(r)$ is virtually identical within
$r_{200}$ for both $\Omega_0=0.2,1$.  For $r\la 0.2r_{200}$ the
virialized portions of the clusters approximately follow the
self-similar profile $\rho\sim r^{-9/4}$ \cite{edbert} and steepen to
$\rho\sim r^{-2.5}$ at larger radii $r\sim (1-1.5)h^{-1}$ Mpc.  When
the infalling mass is included the behavior at small radii is largely
unaffected, but at large radii the $\Omega_0=1$ profile becomes
flatter than $\Omega_0=0.2$: $\rho\sim r^{-2.4}$ for $\Omega_0=1$ and
$\rho\sim r^{-2.5}$ for $\Omega_0=0.2$ over radii $r\sim
(0.5-1)r_{200}$ or $r\sim (0.75-1.5)h^{-1}$ Mpc.  For a given
$\Omega_0$ the profile shapes of clusters of different masses are
essentially identical when expressed in terms of $r_{200}$.

The density and mass profiles appear to agree well with those produced
in the three-dimensional N-body simulations of a standard CDM
cosmology. The $\rho\sim r^{-9/4}$ dependence in the inner regions
giving way to a steeper density profile matches the behavior found by
Anninos \& Norman \shortcite{an} in their detailed study of the effect
of mass resolution on cluster density profiles.  Moreover, the
relative similarity of the density profiles for different masses and
different values of $\Omega_0$ agrees with the conclusions of Navarro
et al. \shortcite{nfw}, aside from the core regions $(r\la
0.2r_{200})$ where Navarro et al. find $\rho\sim r^{-1}$.

\subsection{$\mathbf \Delta M /\, \overline{M}$}

\subsubsection{CDM: $z=0$ $(\lambda_0=0)$}
\label{cdmz0}

\begin{figure}
\centerline{\hspace{0cm}\psfig{figure=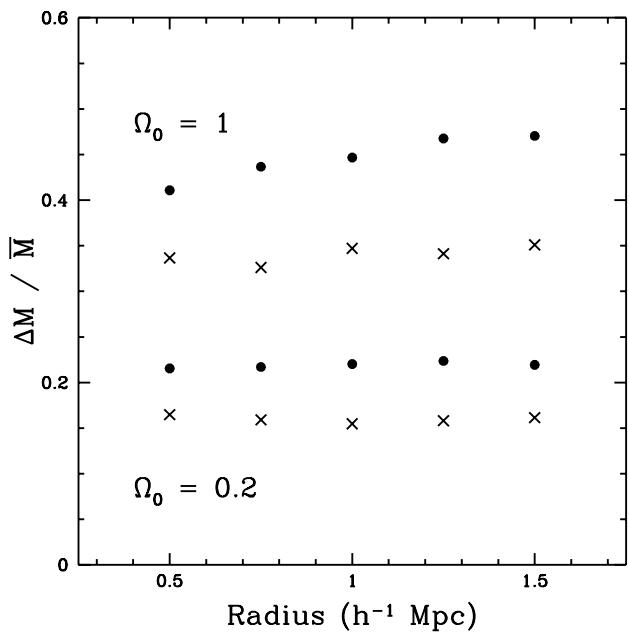,width=0.49\textwidth,angle=0}}
\caption{\label{fig.dmvsr} The radial profile of ${\Delta M
/\,\overline{M}}$ at $z=0$ for CDM models with $\Omega_0=0.2,1$.
Clusters with mass $\approx 3.5\times 10^{14}h^{-1}M_{\sun}$ are shown
as filled circles and those of mass $\approx 1.4\times
10^{15}h^{-1}M_{\sun}$ by crosses. (These masses are approximately the
values within a radius $1.5h^{-1}$ Mpc.)}
\end{figure}

We begin our discussion by examining the results of ${\Delta M
/\,\overline{M}}$ at $z=0$ for CDM models with $\lambda_0=0$. At
present we restrict our discussion to the case where the relaxation
timescale equals $t_{cross}$.  In Figure \ref{fig.dmvsr} we show the
radial dependence of ${\Delta M /\,\overline{M}}$ for a couple of
representative masses with $\Omega_0=0.2,1$. The profiles of ${\Delta
M /\,\overline{M}}$ are nearly constant, though there appears to be a
slight $(\sim 10\%)$ increase with radius for low-mass clusters when
$\Omega_0=1$; i.e. the dynamical state, or degree of virialization, is
not a strong function of radius within $1.5h^{-1}$ Mpc.  The slight
increase of ${\Delta M /\,\overline{M}}$ with radius is reasonable
since the amount of accreting mass becomes fractionally more important
at larger radii, and models with larger $\Omega_0$ have proportionally
more accreting mass\footnote{The similar radial dependence of ${\Delta
M /\,\overline{M}}$ for different masses also indicates that there is
no conceptual advantage to focusing on a scaled radius like
$r_{200}$.}.

To assess the reliability of ${\Delta M/\,\overline{M}}$ with respect
to the simplifications of our model, we compare to the results of the
N-body simulations of BX.  However, BX analyzed the projection of
$\rho^2$ and focused on the mean values of $\log_{10}P_m/P_0$
$(m=2,3)$ of their N-body cluster samples, where
$P_m/P_0\equiv\langle(\Psi^{int}_m)^2\rangle/\langle(\Psi^{int}_0)^2\rangle$
(see \S \ref{mot}). Although these properties prohibit a rigorous
quantitative comparison of ${\Delta M/\,\overline{M}}$ to the BX
results, we expect qualitative similarities. Henceforth we shall
compare our results for ${\Delta M/\,\overline{M}}$ to the $P_m/P_0$
results obtained by BX with these caveats understood.  For our
comparison at hand, the nearly constant radial behavior of ${\Delta M
/\,\overline{M}}$ is similar to $P_m/P_0$, though the $P_m/P_0$
actually show a modest decrease with radius.  This decrease in
$P_m/P_0$ is expected because the properties of the multipole
expansion dictate that higher order moments in the potential decay
much more rapidly than the monopole with increasing distance from a
mass concentration.

\begin{figure}
\centerline{\hspace{0cm}\psfig{figure=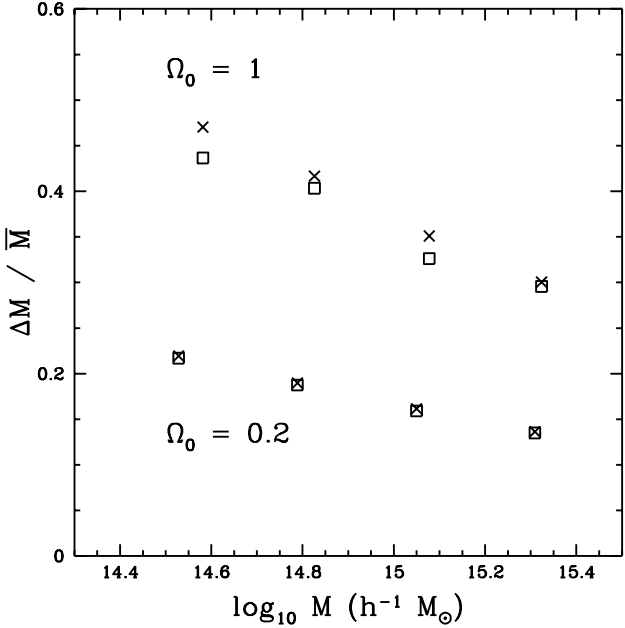,width=0.49\textwidth,angle=0}}
\caption{\label{fig.dmvsm} ${\Delta M /\,\overline{M}}$ evaluated for
radii $r=0.75h^{-1}$ Mpc (boxes) and $r=1.5h^{-1}$ Mpc (crosses) as a
function of total cluster mass within $1.5h^{-1}$ Mpc for
$\Omega_0=0.2,1$ CDM models at $z=0$. }
\end{figure}

In Figure \ref{fig.dmvsm} we display ${\Delta M /\,\overline{M}}$ as a
function of cluster mass (i.e. defined within $1.5h^{-1}$ Mpc).  The
dependence of ${\Delta M /\,\overline{M}}$ on cluster mass is
significant: for a given radius the more massive clusters tend to have
smaller ${\Delta M /\,\overline{M}}$.  This is a reflection of the
shorter relaxation (crossing) times for the more massive clusters
within a given radius.  The slightly steeper dependence for the low
mass clusters within the $1.5h^{-1}$ Mpc aperture just reflects the
slight increase with radius of ${\Delta M /\,\overline{M}}$ for
$\Omega=1$ models discussed above (see Figure \ref{fig.dmvsr}).

Now we focus our attention on how ${\Delta M /\,\overline{M}}$ depends
on $\Omega_0$.  Since observations and N-body simulations give mean
values averaged over a cluster sample, we average ${\Delta M
/\,\overline{M}}$ over the mass function appropriate for our CDM
models.  Although the peaks formalism (BBKS) does not provide a well
defined mass function, the analytic Press-Schechter \cite{ps} mass
function is convenient and a suitable approximation in many cases
(Bond et al. 1991).  For our averaging procedure we compute ${\Delta M
/\,\overline{M}}$ for a few clusters spanning the mass range
$(0.35-3)\times 10^{15}h^{-1}M_{\sun}$ and interpolate ${\Delta M
/\,\overline{M}}$ for arbitrary $M$ over that range.  This function is
then averaged over the Press-Schechter mass function using the
approximations in Viana \& Liddle \shortcite{vl} for CDM models to
yield the mass-averaged fractional accreted mass denoted by
$\langle{\Delta M /\,\overline{M}}\rangle$.

\begin{figure}
\centerline{\hspace{0cm}\psfig{figure=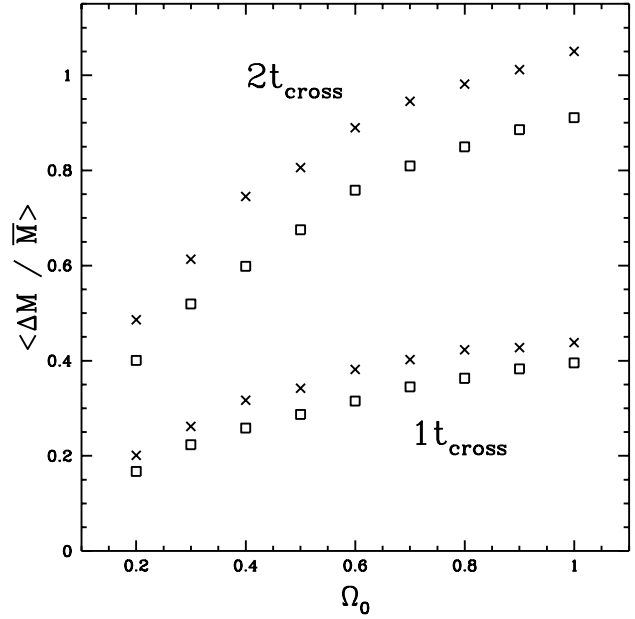,width=0.49\textwidth,angle=0}}
\caption{\label{fig.dmvsom} The mass-averaged fractional accreted mass
evaluated for $r=1h^{-1}$ Mpc at $z=0$ for CDM models as a function of
$\Omega_0$ $(\lambda_0=0)$. The crosses indicate a mass average over
the full range $(0.35-3)\times 10^{15}h^{-1}M_{\sun}$, and the boxes
indicate a lower limit of $7\times 10^{14}h^{-1}M_{\sun}$.  Relaxation
timescales of 1 and 2 crossing times are shown.}
\end{figure}

In Figure \ref{fig.dmvsom} we display $\langle{\Delta M
/\,\overline{M}}\rangle$ within $r=1h^{-1}$ Mpc as a function of
$\Omega_0$ $(\lambda_0=0)$.  As expected, $\langle{\Delta M
/\,\overline{M}}\rangle$ increases with increasing $\Omega_0$:
$\langle{\Delta M /\,\overline{M}}\rangle \cong 0.46\Omega_0^{0.45}$
when averaged over the full mass range $(0.35-3)\times
10^{15}h^{-1}M_{\sun}$, and $\langle{\Delta M /\,\overline{M}}\rangle
\cong 0.40\Omega_0^{0.50}$ when averaged over the smaller range
$(0.7-3)\times 10^{15}h^{-1}M_{\sun}$ corresponding to more massive
clusters.  The slight variation of the exponent of $\Omega_0$ as a
function of mass range is the result of the mass averaging -- we find
that $\Delta M /\,\overline{M}\propto \Omega_0^{0.47}$ applies
essentially over the full mass range $(0.35-3)\times
10^{15}h^{-1}M_{\sun}$.  Given the results of Figures \ref{fig.dmvsr}
and \ref{fig.dmvsm}, it is not surprising that there is little
advantage in going to larger radii as the greatest increase is seen
within $r=1.5h^{-1}$ Mpc for the least massive clusters for which we
obtain the marginally steeper dependence $\langle{\Delta M
/\,\overline{M}}\rangle \cong 0.47\Omega_0^{0.47}$ averaged over the
full mass range.

\begin{figure}
\centerline{\hspace{0cm}\psfig{figure=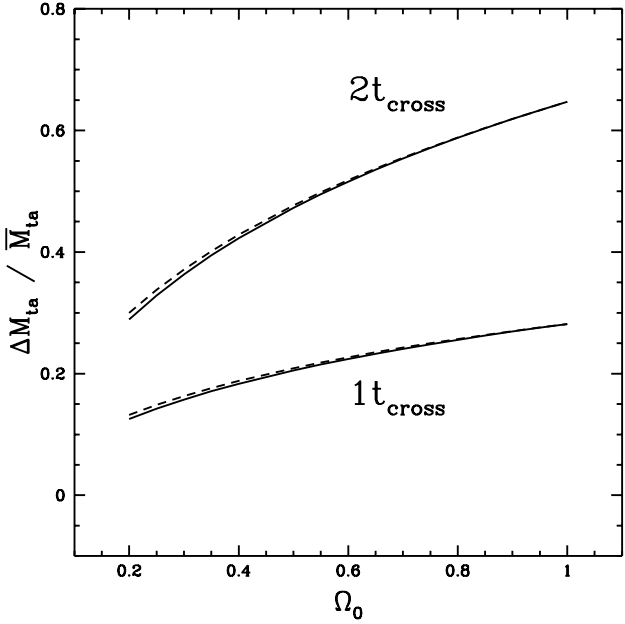,width=0.49\textwidth,angle=0}}
\caption{\label{fig.guess} The fractional ``collapsed'' mass accreted
over the previous relaxation time as a function of $\Omega_0$ for CDM
models at $z=0$.  The collapsed mass is approximately $7\times
10^{14}h^{-1}M_{\sun}$ enclosed within a radius $1.5h^{-1}$ Mpc. The
solid lines indicate the $\lambda_0=0$ models, and dashed lines
indicate $\Omega_0+\lambda_0=1$ models.  Relaxation timescales of 1
and 2 crossing times are shown where $t_{cross}\equiv 0.14/H_0$.}
\end{figure}

To determine the relationship between this $\sim\Omega_0^{1/2}$
dependence and the turn-around times of the mass shells we compare
these results to ${\Delta M_{ta}/\, \overline{M}_{ta}}$.  Setting
$t_{cross}=0.14/H_0$ (see below) we compute ${\Delta M_{ta}/\,
\overline{M}_{ta}}$ as a function of collapsed mass at $z=0$.  In
Figure \ref{fig.guess} we display ${\Delta M_{ta}/\,
\overline{M}_{ta}}$ as a function of $\Omega_0$ for a collapsed mass
$7\times 10^{14}h^{-1}M_{\sun}$.  In agreement with the behavior of
$\Delta M /\,\overline{M}$, we find ${\Delta M_{ta}/\,
\overline{M}_{ta}}\propto \Omega_0^{0.48}$ which is essentially
independent of mass over the full range $(0.35-3)\times
10^{15}h^{-1}M_{\sun}$.  The normalization of ${\Delta M_{ta}/\,
\overline{M}_{ta}}$, however, is $\sim 40\%$ smaller: ${\Delta
M_{ta}/\, \overline{M}_{ta}}\cong 0.28\Omega_0^{0.48}$ versus $\Delta
M /\,\overline{M}\cong 0.43 \Omega_0^{0.47}$ for $7\times
10^{14}h^{-1}M_{\sun}$.  The close agreement between $\Delta M
/\,\overline{M}$ and ${\Delta M_{ta}/\, \overline{M}_{ta}}$ indicates
that the $\sim\Omega_0^{1/2}$ dependence is derived from the
turn-around times of the mass shells. (This dependence is only weakly
affected by the relaxation timescale -- see below.)

We now compare $\langle{\Delta M /\,\overline{M}}\rangle$ to the
N-body results of BX. In order to make the comparison more reasonable,
let us consider $\log_{10} \langle{(\Delta M
/\,\overline{M})^2}\rangle$ within the $0.75h^{-1}$ Mpc aperture, for
which our models give $\langle{(\Delta M
/\,\overline{M})^2}\rangle\cong 0.19\Omega_0^{0.81}$ over the full
mass range. Let us consider the logarithmic difference of this
quantity between a model with $\Omega_0=1$ and $\Omega_0<1$ and denote
this shift by $\Delta$. We can then write, $\Delta =
-0.81\log_{10}\Omega_0$.  Setting $\Omega_0=0.35$ appropriate for BX
(and $c_m=1$, see \S \ref{present}), we obtain $\Delta = 0.4$, in
pretty good agreement with $\Delta\sim 0.5$ obtained by BX considering
that they analyzed $\rho^2$ in projection\footnote{The projection of
the dark matter in clusters appears to yield smaller variations in
$P_m/P_0$ than does the projection of $\rho_{gas}^2$ (see \S 4 of Tsai
\& Buote).}. This qualitative agreement with BX is reassuring on two
accounts: (1) $\langle{\Delta M /\,\overline{M}}\rangle$, which
represents the fractional increase in the monopole of the potential,
clearly does correlate strongly with the next few low order moments;
(2) our simple spherical accretion model is able to reasonably
reproduce the differences in the means of cluster distributions
generated by three-dimensional N-body simulations.

We may also compare $\langle{\Delta M /\,\overline{M}}\rangle$ to the
results of Richstone et al. \shortcite{rlt} who apply the spherical
collapse model to clusters of mass $10^{15}h^{-1}M_{\sun}$ arising
from homogeneous density perturbations, which are assumed to form at
the time equal to twice the turn-around time of the perturbation.
Richstone et al. compute the quantity $\delta F$, the fraction of
present-day clusters which formed within the last time interval
$\delta t$, as a function of $\Omega_0$ and $\lambda_0$. Consulting
Figure 3 of Richstone et al. we see that $\delta F\sim\Omega_0$ for
substructure survival time $\delta t=0.1/H_0$ (a timescale similar to
our $1t_{cross}$).  (A similar dependence of $\delta F$ on $\Omega_0$
is seen in Figure 13 of Lacey \& Cole 1993.)

Although $\delta F$ is computed from the collapse times of density
perturbations, it depends on these collapse times through the error
function; i.e. $\langle{\Delta M /\,\overline{M}}\rangle$ and $\delta
F$ are qualitatively different in how they depend on the collapse
times of shells and thus $\Omega_0$.  In fact, we can repeat the
exercise above for computing the mean shift between models with
$\Omega_0=1$ and $\Omega_0<1$ by replacing $\langle{\Delta M
/\,\overline{M}}\rangle$ with $\delta F$ in which case we obtain
$\Delta = -2\log_{10}\Omega_0$. For $\Omega_0=0.35$, $\Delta=0.9$
which is almost twice the mean shift found by BX.  The significantly
better agreement of the $\Delta$ predicted by $\langle{\Delta M
/\,\overline{M}}\rangle$ over $\delta F$ with the results of BX
demonstrates that a much closer relationship exists between
$\langle{\Delta M /\,\overline{M}}\rangle$ and $P_m/P_0$ than between
$\delta F$ and $P_m/P_0$ as would be expected.

The crossing times for the ($\lambda_0=0$) CDM models are not a strong
function of mass or $\Omega_0$: $t_{cross}=(0.12-0.17)/H_0$ with a
typical value $t_{cross}\cong 0.14/H_0\cong 1.4\times 10^{9}h^{-1}$
yr. (The largest value corresponds to a mass $3.5\times
10^{14}h^{-1}M_{\sun}$ for $\Omega_0=1$.)  We show in Figure
\ref{fig.dmvsom} the results for $\langle{\Delta M
/\,\overline{M}}\rangle$ for the case where the relaxation timescale
is set to $2t_{cross}$. Although the values of $\langle{\Delta M
/\,\overline{M}}\rangle$ have approximately doubled, the power-law
dependence on $\Omega_0$ remains virtually unchanged: $\langle{\Delta
M /\,\overline{M}}\rangle \cong 1.08\Omega_0^{0.45}$ when averaged
over the full mass range $(0.35-3.5)\times 10^{15}h^{-1}M_{\sun}$, and
$\langle{\Delta M /\,\overline{M}}\rangle \cong 0.94\Omega_0^{0.49}$
when averaged over the smaller range $(0.7-3)\times
10^{15}h^{-1}M_{\sun}$.  This behavior is mirrored by ${\Delta
M_{ta}/\, \overline{M}_{ta}}$ which maintains its $\Omega_0^{0.48}$
dependence for $2t_{cross}$ (Figure \ref{fig.guess}). (Note this same
type of $\Omega_0$-dependence also applies for a relaxation timescale
of $0.5t_{cross}$.)

A noticeable weakening of the dependence on $\Omega_0$ occurs when the
relaxation timescale is increased to $3t_{cross}$: $\langle{\Delta M
/\,\overline{M}}\rangle \cong 1.71\Omega_0^{0.35}$ when averaged over
the full mass range $(0.35-3.5)\times 10^{15}h^{-1}M_{\sun}$, and
$\langle{\Delta M /\,\overline{M}}\rangle \cong 1.61\Omega_0^{0.45}$
when averaged over the smaller range $(0.7-3)\times
10^{15}h^{-1}M_{\sun}$.  For a cluster that accretes an increasingly
larger fraction of its current mass during the relaxation time,
$\Delta M /\,\overline{M}$ approaches a value of 2 independent of
$\Omega_0$.  Because the smaller masses tend to have larger $\Delta M
/\,\overline{M}$, their dependence on $\Omega_0$ weakens more rapidly
with redshift than the more massive clusters.

${\Delta M_{ta}/\, \overline{M}_{ta}}$ is slower to respond to this
effect. For $3t_{cross}$, ${\Delta M_{ta}/\, \overline{M}_{ta}}\cong
1.21\Omega_0^{0.46}$ for a mass $3.5\times
10^{14}h^{-1}M_{\sun}$. This becomes $1.80\Omega_0^{0.39}$ at
$4t_{cross}$ and approximately $2.09\Omega_0^{0.17}$ at
$5t_{cross}$. The departure of ${\Delta M_{ta}/\, \overline{M}_{ta}}$
from $\Delta M /\,\overline{M}$ indicates that the dynamical evolution
depends more on the virialized structure of the cluster for large
relaxation times, or equivalently for larger redshifts. (This effect
is discussed in more detail in \S \ref{cdmz}.)

For our current purposes the trend of $\langle{\Delta M
/\,\overline{M}}\rangle$ with $\Omega_0$ is of principal importance,
not the actual value\footnote{This trend is quite insensitive to the
slope of the power spectrum -- see end of \S \ref{pl}.}.  Recall that
in this paper we do not intend to compare $\langle{\Delta M
/\,\overline{M}}\rangle$ directly to observations (\S \ref{present}),
but rather we expect the next few moment ratios of the gravitational
potential, which can be observed, to be strongly correlated with
$\langle{\Delta M /\,\overline{M}}\rangle$ (see \S \ref{mot}).  Since
1-2 crossing times should effectively cover the reasonable range of
relaxation timescales for potential fluctuations characterized by the
lower order moments, the utility of $\langle{\Delta M
/\,\overline{M}}\rangle$ for studying the dependence of cluster
morphology is hardly affected by the uncertainty in the relaxation
timescale. This behavior contrasts with that of $\delta F$ \cite{rlt}
and related measures \cite{lc} whose value is to be directly compared
to the ``frequency of substructure''.

Changing the value of $\sigma_8$ (equation \ref{eqn.sigma8}) has
virtually no effect on $\Delta M /\,\overline{M}$. For
$\sigma_8=0.52,1$ $(\Omega_0=1)$ we find that ${\Delta M
/\,\overline{M}}$ varies by $\la 1\%$.  This similarity occurs because
the amplitude of $P(k)_{clus}$ arising from a higher $\sigma_8$ is
mostly compensated for by a reduced cluster bias factor, $b_{clus}$.
This insensitivity to $\sigma_8=0.5-1$ is also found in the mean
$\log_{10}P_m/P_0$ distributions of BX.

Finally, we investigated the effect of using $\Omega_B$ appropriate
for the gas fractions in clusters \cite{wf}. In our models this only
has the effect of slightly changing $H_0$ determined for a particular
value of $\Omega_0$ (see definition of $\Gamma$ in \S
\ref{cospar}). As a result, $\langle{\Delta M /\,\overline{M}}\rangle$
is essentially unaffected -- we find results in excellent agreement
with those obtained above with the BBN $\Omega_B$. Of course, it would
be better to test the effects of a greater baryon fraction by
incorporating the gravitational dynamics of a dissipational component
into the model.

\subsubsection{CDM: $\Omega_0+\lambda_0=1$ at $z=0$}
\label{lambdacdm}

\begin{figure}
\centerline{\hspace{0cm}\psfig{figure=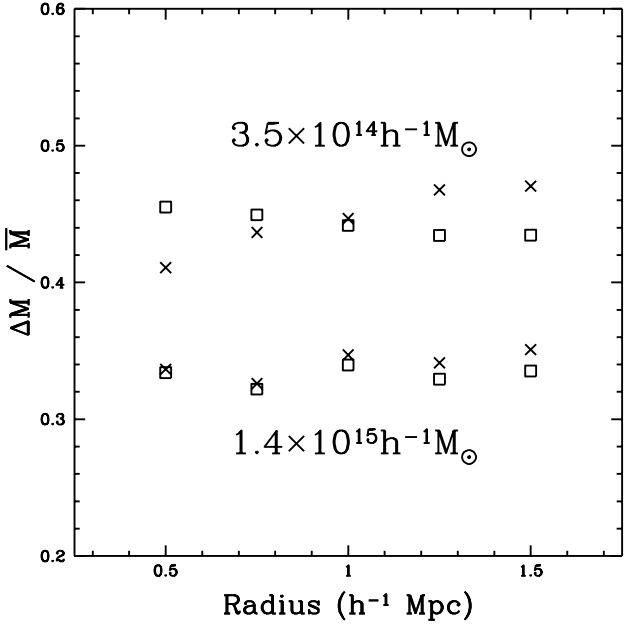,width=0.49\textwidth,angle=0}}
\caption{\label{fig.dmvsrlam} $\Delta M /\,\overline{M}$ versus radius
for models with $\Omega_0=0.2$, $\lambda_0=0.8$ (boxes) and
$\Omega_0=1$ (crosses).}
\end{figure}

\begin{figure}
\centerline{\hspace{0cm}\psfig{figure=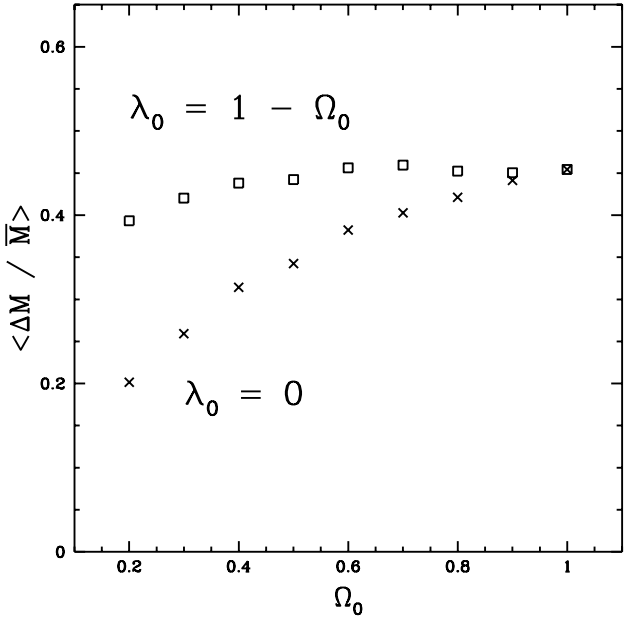,width=0.49\textwidth,angle=0}}
\caption{\label{fig.dmvsomlam} $\langle{\Delta M
/\,\overline{M}}\rangle$ versus $\Omega_0$ for $\Omega_0+\lambda_0=1$
models (boxes) and $\lambda_0=0$ models (crosses). The relaxation
timescale is $1t_{cross}$.}
\end{figure}

We now turn our attention to the zero-curvature models,
$\Omega_0+\lambda_0=1$.  In a low-density universe with zero-curvature
only recently has the effect of a cosmological constant significantly
imprinted itself on the cosmic dynamics; i.e. the transition from
$\Omega_0\cong 1$ to $\Omega_0<1$ occurs considerably later and much
more rapidly than in the $\lambda_0=0$ case.  Hence, although the
shells which collapsed at the present epoch turned around at times
similar to those in $\lambda_0=0$ models, the previously collapsed
shells turned around with times governed by a universe with
$\Omega_0\rightarrow 1$.

As a result, the density profiles of the zero-curvature models have
essentially the same slope as the $\Omega_0=1$ models over radii $\sim
(0.5-1)r_{200}$ (see \S \ref{masspro}).  However, the profile of the
zero-curvature models is flatter at smaller radii, $r\la 0.2r_{200}$,
indicating that the mean density within the central regions of the
clusters is considerably smaller in zero-curvature models, consistent
with the arguments given in \S 4 of Richstone et
al. \shortcite{rlt}. Because of this decrease in density, the
half-mass radii are larger, the velocity dispersions smaller, and thus
the crossing times (equation \ref{eqn.tcross}) are considerably larger
in the low-density zero-curvature models than in their $\lambda_0=0$
counterparts: $t_{cross}=(0.20-0.23)/H_0$ over the full mass range for
the $\Omega_0=0.2$ and $\lambda_0=0.8$ model, a factor of $\sim 1.7$
larger than the open $\Omega_0=0.2$ model.  In fact, with respect to
$\Omega_0=1$, $t_{cross}$ is $\sim 40\%$ {\it larger} for the
zero-curvature $\Omega_0=0.2$ model but $\sim 15\%$ {\it smaller} for
the open $\Omega_0=0.2$ model.  This considerably stronger and
qualitatively different dependence of $\Delta M /\,\overline{M}$ on
crossing time as a function of $\Omega_0$ causes the low-density
zero-curvature models to have qualitatively different behavior than
their $\lambda_0=0$ counterparts.

We display in Figure \ref{fig.dmvsrlam} the radial profile of $\Delta
M /\,\overline{M}$ for the $\Omega_0=0.2$, $\lambda_0=0.8$ model and
the $\Omega_0=1$ model for a couple of representative masses. For
$M\ga 7\times 10^{14}h^{-1}M_{\sun}$, the profile of the
zero-curvature model is essentially flat but almost of the same
magnitude as the $\Omega_0=1$ model. As expected, the differences
between the models becomes most noticeable at larger radii because the
accretion of the most recent shells is suppressed in the zero-curvature
model; i.e. the recently collapsed shells approximately behave as in
the $\Omega_0=0.2$, $\lambda_0=0$ case.

The $\Omega_0=0.2$ zero-curvature radial profile of $\Delta M
/\,\overline{M}$ slightly decreases with radius for $M= 3.5\times
10^{14}h^{-1}M_{\sun}$ and actually exceeds the $\Omega_0=1$ values
for $r<1h^{-1}$ Mpc.  This decreasing profile indicates that shells
with turn-around radii $r_m\la 2h^{-1}$ Mpc dominate the contribution
to $\Delta M /\,\overline{M}$ over the previous crossing time -- since
the crossing time is significantly larger than for $\Omega_0=1$, the
zero-curvature model accretes shells at higher redshift which
turned-around at smaller radii.

In Figure \ref{fig.dmvsomlam} we display $\langle{\Delta M
/\,\overline{M}}\rangle$ versus $\Omega_0$ for the zero-curvature
models. In order to accentuate the dependence on $\Omega_0$ we focus
on the largest aperture considered, $r=1.5h^{-1}$ Mpc. We find that
these models behave very nearly as $\Omega_0=1$ and depend only weakly
on $\Omega_0$: $\langle{\Delta M /\,\overline{M}}\rangle\sim
\Omega_0^{0.1}$.  This behavior is qualitatively consistent with the
linear growth of density fluctuations: the ratio of linear growth
factors for the $\Omega_0=0.2$, $\lambda_0=0.8$ model to $\Omega_0=1$
is 0.70 compared to a ratio of 0.33 for the open $\Omega_0=0.2$ model
to $\Omega_0=1$. That is, from linear growth we would expect the
amount of accreted mass in clusters in the $\Omega_0=0.2$
zero-curvature model to be more similar to the $\Omega_0=1$ model than
to the open $\Omega_0=0.2$ model (e.g. Peebles 1984).  However, the
N-body simulations of BX indicate that the means of the
$\log_{10}P_m/P_0$ distributions of their zero-curvature models,
though intermediate with the $\Omega_0=0.35,1$ cases they examined,
nevertheless leaned more closely to the $\Omega_0=0.35$ open model for
radii $(0.4-1.2)h^{-1}$ Mpc.

We emphasize that this weak $\Omega_0$-dependence arises from the
increased crossing times for the low-density zero-curvature models. In
Figure \ref{fig.guess} we display $\Delta M_{ta}\, /\overline{M}_{ta}$
for the zero-curvature models with $t_{cross}=0.14/H_0$ appropriate
for the $\lambda_0=0$ models. In this case $\Delta M_{ta}\,
/\overline{M}_{ta}\propto \Omega_0^{0.45}$ which is only slightly
weaker than the $\Omega_0$-dependence for the $\lambda_0=0$
models. This $\sim\Omega_0^{0.45}$ dependence applies to
$\langle{\Delta M /\,\overline{M}}\rangle$ if we fix
$t_{cross}=0.14/H_0$ for the zero-curvature models.  (This behavior is
also found for $\delta F$ by Richstone et al. who also used a constant
relaxation timescale for all models.)

Our procedure of setting the relaxation timescale equal to the
crossing time is only an approximation.  Because the crossing times in
the zero-curvature models vary by almost a factor of 3 more than in
the open models over $\Omega_0=0.2-1$, the accuracy associated with
this procedure is considerably more important for studying the
$\Omega_0$-dependence of $\Delta M\, /\overline{M}$ in the
zero-curvature models.

In addition, other factors contribute to a computational
uncertainty. Because the increase in crossing time is due primarily to
the flatter density profile at small radii, the error due to the
number of shells becomes more important. Also, issues like the
smoothing length of the power spectrum become more important at radii
$\la 0.1r_{200}$. Hence, the detailed features in the profile of
$\langle{\Delta M /\,\overline{M}}\rangle$ vs $\Omega_0$ in Figure
\ref{fig.dmvsomlam} (e.g. the $\Omega_0\approx 0.7$ values exceeding
$\Omega_0=1$) are probably due to the uncertainty in the relaxation
timescale.  Further study of the relaxation timescale for the
zero-curvature models using three dimensional N-body simulations would
certainly be of value here.

Nevertheless, it is clear that the low-density zero-curvature models
do not mimic the corresponding open models and thus cluster
morphologies do indeed distinguish between these two cases at $z=0$
\footnote{BX found that the means of the $\log_{10}P_m/P_0$
distributions of 39 simulated clusters for their $\Omega_0=0.35$ open
and zero-curvature models, though systematically larger for the
zero-curvature models as indicated by the non-overlapping error bars
derived from bootstrap re-sampling, were formally consistent in terms
of the Student's-t test. It appears that a larger cluster sample is
required to formally differentiate the means of the two models. This
has been confirmed by Thomas et al. \shortcite{thomas}.}. We find that
the difference between the models arises from the detailed virialized
structure of the clusters (i.e. crossing times), which is consistent
with the expectations of Lahav et al. \shortcite{lahav} that a
cosmological constant, ``could only have changed the density profile
by its effect on the non-linear motion of shells.''

\subsubsection{CDM: $z>0$}
\label{cdmz}

\begin{figure*}
\parbox{0.49\textwidth}{
\centerline{\psfig{figure=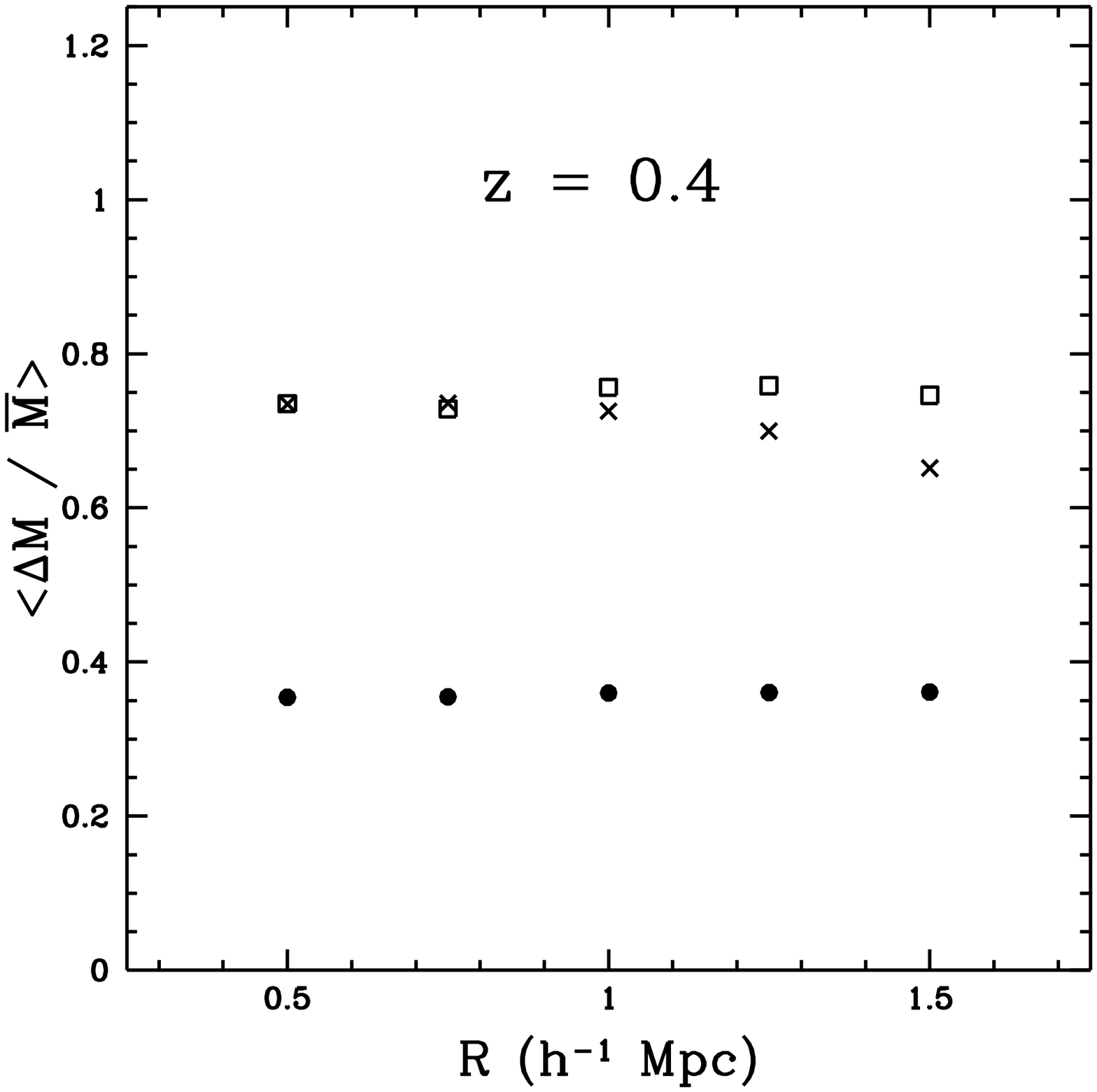,angle=0,height=0.3\textheight}}
}
\parbox{0.49\textwidth}{
\centerline{\psfig{figure=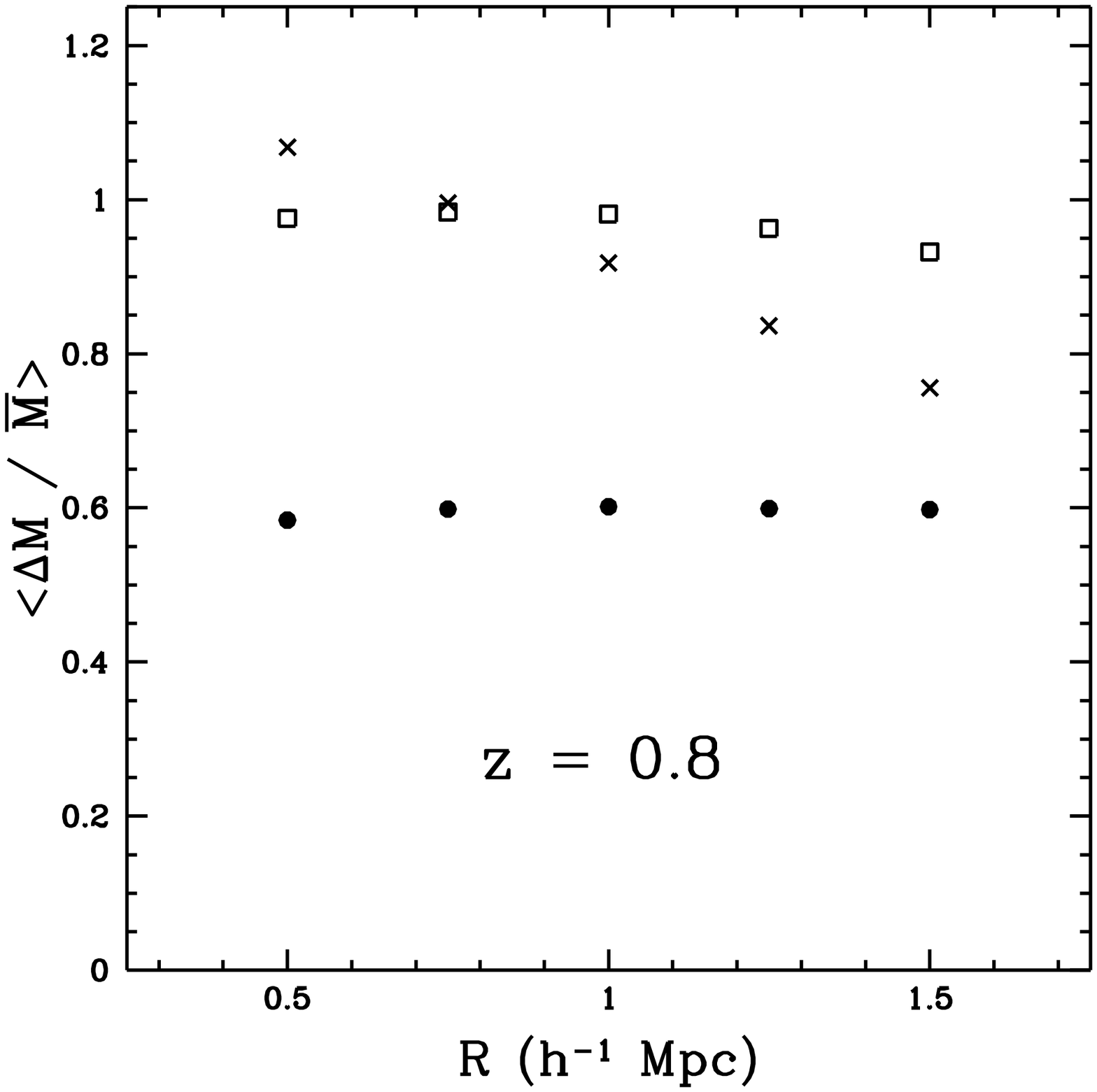,angle=0,height=0.3\textheight}}
}
\caption{\label{fig.dmrz} The radial profile of $\langle{\Delta M
/\,\overline{M}}\rangle$ as a function of redshift for CDM models
$\Omega_0=0.2$, $\lambda_0=0$ (filled circles), $\Omega_0=0.2$,
$\lambda_0=0.8$ (boxes), and $\Omega_0=1$ (crosses). See text for
details of the mass average.}
\end{figure*}

Let us now consider the evolution of $\Delta M /\,\overline{M}$ with
redshift in the context of the CDM model.  For compactness we restrict
our discussion to a relaxation timescale of $1t_{cross}$ and
$\Omega_B$ equal to the BBN value. (The effect of different relaxation
timescales is mentioned at the end of this section.)  Since the
cluster population becomes less massive with increasing $z$, we
analyze $\Delta M/\,\overline{M}$ averaged over masses
$[(0.7-3)/(1+z)]\times 10^{15}h^{-1}M_{sun}$, where the $(1+z)$ factor
is intended to account for the reduction in mass according to the
self-similar accretion law \cite{edbert}.  We focus on a slightly
higher mass range because a feasible observational sample either from
lensing or X-rays will be biased towards more massive clusters. The
qualitative results presented in this section remain if the lower
limit on the mass average is halved.

In Figure \ref{fig.dmrz} we display the radial profile of
$\langle{\Delta M/\,\overline{M}}\rangle$ at $z=0.4,0.8$ for models
with $\Omega_0=0.2,1$ $(\lambda_0=0)$ and $\Omega_0=0.2,
\lambda_0=0.8$. Over redshifts $z\sim 0-0.4$ the behavior of
$\langle{\Delta M/\,\overline{M}}\rangle$ with radius does not change
appreciably.  The shape of the open $\Omega_0=0.2$ profile remains
constant over this interval, but the $\Omega_0=1$ profile shifts from
being slightly increasing with radius (Figure \ref{fig.dmvsr}) to
slightly decreasing.  As discussed in \S \ref{lambdacdm} for the
zero-curvature models, the decreasing profile indicates that as $z$
increases the clusters in the $\Omega_0=1$ models are growing faster
in their inner regions; i.e. the accreted mass is dominated by shells
with turn-around radii $\la 2h^{-1}$ Mpc. This contrasts with the
situation at $z=0$ for the $\lambda_0=0$ models where the accretion of
infalling shells contributes almost entirely to large radii, which is
consistent with the behavior of the exact self-similar solution
\cite{edbert}.

The shape of the fractional accreted mass profile of the
zero-curvature $\Omega_0=0.2$ model is similar to that of the
corresponding open model, but the normalization is similar to
$\Omega_0=1$ as we found for the $z=0$ case (with the same caveats
regarding the crossing time effects discussed in \S \ref{lambdacdm}).
The profile of the zero-curvature $\Omega_0=0.2$ model does not
decrease like the $\Omega_0=1$ model because the shells which affect
most the outer radii are still determined by $\Omega_0\approx
0.2$. That is, for the zero-curvature models the redshift indicating
when the term involving the cosmological constant in the Friedmann
equation dominates the matter term is given by, $(1+z_{trans})^3 =
\Omega_0^{-1}-1$, after which linear growth is suppressed (e.g. \S 13
Peebles 1980). For $\Omega_0=0.2$, we have $z_{trans}=0.6$.

Indeed, proceeding to higher redshift, $z=0.8$, we see in Figure
\ref{fig.dmrz} that the profile of the zero-curvature $\Omega_0=0.2$
model now decreases with radius indicating that the matter term is
beginning to dominate the Friedmann equation.  The $\langle{\Delta
M/\,\overline{M}}\rangle$ radial profile for $\Omega_0=1$ steepens
while that for the open $\Omega_0=0.2$ model remains essentially flat.
For redshifts $\ga 1$ the profiles of the zero-curvature
$\Omega_0=0.2$ model and $\Omega_0=1$ model have very similar
shapes. Even the open model starts to steepen, though from the
argument from linear theory we would not expect the transition to
occur until approximately $z_{trans}=3$.  We emphasize, however, that
the precise redshifts of these occurrences depend on the relaxation
timescale. For longer relaxation timescales, these properties occur at
lower redshifts and vice versa.

\begin{figure}
\centerline{\hspace{0cm}\psfig{figure=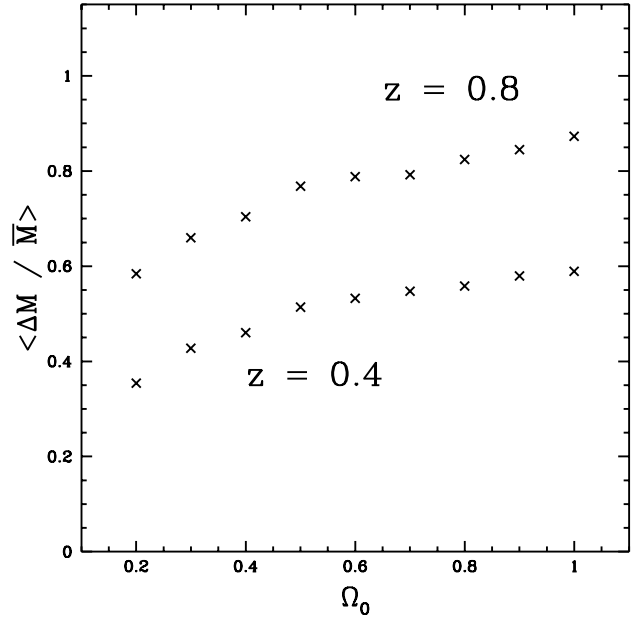,width=0.49\textwidth,angle=0}}
\caption{\label{fig.dmvsomz} $\langle{\Delta M
/\,\overline{M}}\rangle$ evaluated for $r=0.5h^{-1}$ Mpc at
$z=0.4,0.8$ for CDM models as a function of $\Omega_0$
$(\lambda_0=0)$. See text for details of the mass averaging.}
\end{figure}

For the higher redshifts it is clear that the smaller aperture sizes
optimize differences in $\Omega_0$ $(\lambda_0=0)$.  In Figure
\ref{fig.dmvsomz} we show $\langle{\Delta M /\,\overline{M}}\rangle$
computed within a radius $0.5h^{-1}$ Mpc as a function of $\Omega_0$
for $z=0.4,0.8$.  Although the value of $\langle{\Delta M
/\,\overline{M}}\rangle$ increases with increasing redshift indicating
that the cluster morphologies are more disturbed, its sensitivity to
$\Omega_0$ decreases significantly with redshift: $\langle{\Delta M
/\,\overline{M}}\rangle\approx 0.60\Omega_0^{0.29}$ for $z=0.4$ and
$\langle{\Delta M /\,\overline{M}}\rangle\approx 0.87\Omega_0^{0.23}$
for $z=0.8$. (For comparison, for $r=1.5h^{-1}$ Mpc we obtain
$\langle{\Delta M /\,\overline{M}}\rangle\approx 0.67\Omega_0^{0.22}$
for $z=0.4$ and $\langle{\Delta M /\,\overline{M}}\rangle\approx
0.75\Omega_0^{0.07}$ for $z=0.8$.) The weakening trend with redshift
is even observed as early as $z=0.2$ where $\langle{\Delta M
/\,\overline{M}}\rangle\approx 0.52\Omega_0^{0.37}$ in the $1h^{-1}$
Mpc aperture.

This decrease in the sensitivity to $\Omega_0$ arises because of two
effects. First, as $z$ increases the dynamics of the universe
approaches that of the Einstein-de Sitter case regardless of the
present value of $\Omega_0$ and $\lambda_0$. Second, the relaxation
timescale becomes an increasingly larger fraction of the age of the
universe, and thus $\langle{\Delta M
/\,\overline{M}}\rangle\rightarrow 2$ regardless of $\Omega_0$ and
$\lambda_0$.

However, the decline in sensitivity to $\Omega_0$ is more rapid for
$\Delta M /\,\overline{M}$ than for ${\Delta M_{ta}/\,
\overline{M}_{ta}}$.  We find that the relationship ${\Delta M_{ta}/\,
\overline{M}_{ta}}\sim\Omega_0^{1/2}$ holds from $z=0$ until $z\approx
2$ at which point ${\Delta M_{ta}/\, \overline{M}_{ta}}\sim 2$ and the
$\Omega_0$-dependence vanishes.  Since ${\Delta M_{ta}/\,
\overline{M}_{ta}}$ depends only on the turn-around times of shells
and not the detailed virialized cluster structure, its scales more
similarly to what is expected from linear theory: as remarked above,
for the open $\Omega_0=0.2$ model the transition to the Einstein-de
Sitter phase occurs approximately at $z_{trans}=3$.  However, this
redshift is not reached because the crossing time is essentially the
age of the universe at $z\approx 2$. Clearly, taking into account the
detailed virialized structure of the clusters is important for higher
redshifts.

\subsubsection{Power law}
\label{pl}

\begin{figure}
\centerline{\hspace{0cm}\psfig{figure=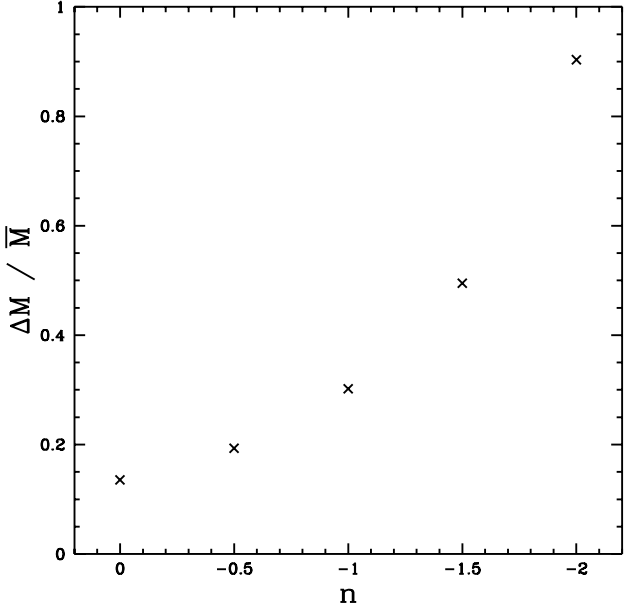,width=0.49\textwidth,angle=0}}
\caption{\label{fig.dmvsn} $\Delta M /\,\overline{M}$ for cluster of
mass $7\times 10^{14}h^{-1}M_{\sun}$ evaluated for $r=1h^{-1}$ Mpc at
$z=0$ for power-law models with $\Omega_0=1$ as a function of spectral
index $n$.}
\end{figure}

Now we switch our focus to models having power-law power spectra.  In
Figure \ref{fig.dmvsn} we display ${\Delta M /\,\overline{M}}$ vs
spectral index $n$ at $z=0$ for clusters of mass $7\times
10^{14}h^{-1}M_{\sun}$; other masses have very similar profiles.  For
the moment we concentrate on models with $\Omega_0=1$ and have a
relaxation time of $1t_{cross}$.  The value of ${\Delta M
/\,\overline{M}}$ for $n\approx -1.3$ agrees with the $\Omega_0=1$ CDM
value as expected for clusters.  The fractional accreted mass
increases as $n$ decreases and varies more over the range $n=0,-2$
($\Omega_0=1$) than it does as a function of $\Omega_0$ for CDM models
over the range $\Omega_0=0.2,1$; ${\Delta M/\,\overline{M}}\approx
0.13(-n)^{5/2}+0.15.$

This $n$-dependence may be understood in terms of the peaks formalism
as follows.  Peaks of a given height in $n=0$ models are much more
isolated than peaks of similar height in $n=-2$ models; i.e. a cluster
of mass $\sim 10^{15}h^{-1}M_{\sun}$ is an exceedingly rare high peak
$\nu\sim 50$ in $n=0$ models whereas such a cluster is a ``normal''
$\nu\sim 3$ peak for $n=-2$ models.  Models with smaller $n$ have more
power on small scales and hence more possible mass to accrete at later
times, which is to be regulated by the value of $\Omega_0$. Thus,
regardless of the value of $\Omega_0=0.2-1$, the huge peak in the
$n=0$ model has a smaller value of ${\Delta M /\,\overline{M}}$ than
the normal peak in $n=-2$ model.

However, real clusters are not $\nu\sim 50$ peaks. Here we experience
the principal flaw in the peaks formalism in that merging is not taken
into account (although see Bond \& Myers 1996), which particularly
affects models with $n\ge -1$ because in this case the collapse
dynamics of a peak is not dominated by its initial density
distribution \cite{bern}.  We have argued (see \S \ref{mot}) that the
effects of merging should not affect the mean of ${\Delta M
/\,\overline{M}}$ when averaged over a cluster sample which is most
relevant for studying $\Omega_0$.  Rather, the fluctuations induced in
${\Delta M/\,\overline{M}}$ by mergers necessarily affects the
variance.

To understand this qualitatively let us envision clusters formed in
$n=0,-2$ models considering the effects of mergers. In the $n=0$
model, since there is little mass on smaller scales for a cluster of
reasonable peak height to accrete, ${\Delta M /\,\overline{M}}$ will
tend to be small for many clusters. However, when a cluster merges
with another of comparable size (it will be similar in size since
there is little small-scale mass) ${\Delta M /\,\overline{M}}$ will be
comparatively large. In comparison, clusters in the $n=-2$ model have
significant small-scale mass to accrete so ${\Delta M
/\,\overline{M}}$ will not vary as much from cluster to cluster as in
the $n=0$ model; i.e. when averaged over a sample of $n=0$ clusters,
${\Delta M /\,\overline{M}}$ will have a broader distribution than in
the $n=-2$ model.

The three dimensional N-body simulations of BX support this
qualitative picture: the distributions of $\log_{10}P_m/P_0$ for the
$n=0$ model indeed have significantly larger variances than for
$n=-2$. (But the means of the distributions are largely the same.)
Hence, for studying the effect of the slope of the power spectrum on
the relationship between ${\Delta M /\,\overline{M}}$ and $\Omega_0$
merging must be properly taken into account.  Since our spherical
model does not account for the effects of mergers, it is imperative to
keep the shape of the power spectrum approximately fixed when
comparing models with different $\Omega_0$ as we have done in the
previous sections.  It is also imperative to insure that the effective
$n$ on cluster scales is less than $-1$ so that the collapse dynamics
is determined mostly by the initial peak density distribution
\cite{bern}.  

\begin{figure}
\centerline{\hspace{0cm}\psfig{figure=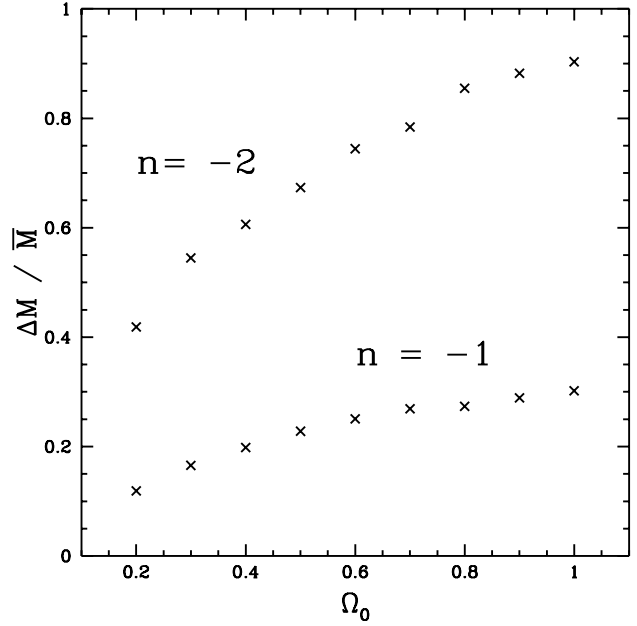,width=0.49\textwidth,angle=0}}
\caption{\label{fig.dmvsnom} $\Delta M /\,\overline{M}$ vs $\Omega_0$
for power-law models with $n=-1,-2$ for clusters of mass $7\times
10^{14}h^{-1}M_{\sun}$ evaluated for $r=1h^{-1}$ Mpc at $z=0$.}
\end{figure}

Although it is important when comparing $\Delta M/\,\overline{M}$ for
models with different $\Omega_0$ to insure that $n$ is approximately
fixed, the value of $n$ is not important. In Figure \ref{fig.dmvsnom}
we show $\Delta M /\,\overline{M}$ as a function of $\Omega_0$ for
power-law models with $n=-1,-2$. Although the values of the $n=-2$
models are approximately double those of the $n=-1$ models, the
dependence on $\Omega_0$ is similar in each: $\Delta
M/\,\overline{M}\approx 0.31\Omega_0^{0.51}$ for $n=-1$ and $\Delta
M/\,\overline{M}\approx 0.93\Omega_0^{0.46}$ for $n=-2$.

\section{Discussion and conclusions}
\label{disc}

We have constructed a simple, intuitive model to study the dependence
of substructure in clusters on $\Omega_0$, $\lambda_0$, and $z$.  We
characterize the importance of substructure and non-ellipsoidal
features in the mass of a cluster, or equivalently the dynamical
state, in terms of, ${\Delta M /\,\overline{M}}$, the fractional
amount of mass accreted over the previous relaxation time within a
radius $r$, where $\Delta M$ is the mass increase and $\overline{M}$
is the average mass within $r$ over the relaxation time.  This
fractional accreted mass is the monopole term in the multipole
expansion of the fractional increase in the gravitational potential
due to matter interior to $r$. Since, by hypothesis, ${\Delta M
/\,\overline{M}}$ determines the importance of substructure and
non-ellipsoidal features, it must strongly correlate with the next few
multipole terms which are approximately the ratios,
$\Phi^{int}_l/\Phi^{int}_0$.

These multipole ratios, or rather their projected counterparts,
$P_m/P_0\equiv\langle(\Psi^{int}_m)^2\rangle/\langle(\Psi^{int}_0)^2\rangle$,
are directly observable and share the same type of correlation with
${\Delta M /\,\overline{M}}$ (see \S \ref{obs}). With N-body
simulations the correlations between ${\Delta M /\,\overline{M}}$ and
$P_m/P_0$ can be explicitly quantified and thus allow a measurement of
$\Omega_0$ to be obtained directly from cluster observations with
gravitational lensing and X-rays.  In this paper we have examined the
differences of ${\Delta M /\,\overline{M}}$ in models with different
$\Omega_0$ (and $P(k)$) which are related to the differences in
$P_m/P_0$ without requiring knowledge of the actual correlation.  (The
reasonable agreement found for the mean shift between $\Omega_0=1$ and
$\Omega_0=0.35$ for the $P_m/P_0$ of the N-body clusters in BX and
${\Delta M /\,\overline{M}}$ provides quantitative justification for
this procedure -- see \S \ref{cdmz0}.)  That is, we predict how the
low order $P_m/P_0$ should behave with $\Omega_0$, $\lambda_0$, and
$z$. In this way we can acquire intuition of how clusters evolve
dynamically in different cosmologies, and, as we discuss below, we can
identify the best ways to constrain $\Omega_0$, $\lambda_0$ with much
greater ease than via computationally expensive three dimensional
N-body simulations.  We compute ${\Delta M /\,\overline{M}}$ using the
spherical accretion model in the adiabatic approximation of Ryden \&
Gunn \shortcite{rg}.

Our model extends and improves upon previous studies in several
respects. First, we compute complete one-dimensional mass
distributions of clusters using the spherical accretion model, where
the initial density profiles of the clusters are given by the peaks
formalism (BBKS). The relaxation time, taken to be a multiple of the
crossing time, is computed individually for each cluster.  In
contrast, previous studies (Richstone et al. 1992; Kauffmann \& White
1993; Lacey \& Cole 1993) did not compute mass distributions of
individual clusters and assumed a constant relaxation timescale for
all clusters. Moreover, Richstone et al. assumed clusters arise from
homogeneous density fluctuations on one particular mass scale.

The principal difference, however, between our work and the previous
studies mentioned above lies in how they relate to observations.  The
previous studies predict the fraction of clusters that are expected to
show evidence of substructure. In terms of observations this is
generally referred to as the ``frequency of substructure'' and
interpreted as the fraction of observed clusters that exhibit multiple
density peaks.  Because these previous theoretical studies do not
indicate the type of substructure or morphological features to be seen
in clusters, it is unclear how to compare their predictions with
observations; i.e. what sizes of multiple density peaks are required
before a cluster is to be considered unformed?  This ambiguity is
resolved in our model because ${\Delta M /\,\overline{M}}$ is expected
to correlate strongly with $\Phi^{int}_l/\Phi^{int}_0$ for small $l$
and with $\Psi^{int}_m/\Psi^{int}_0$ for small $m$. These multipole
ratios quantify the cluster morphology explicitly.

Because the previous studies do not specify the type of substructure,
the relaxation timescale (or rather the survival time of substructure)
is very uncertain -- with some studies indicating uncertainty up to
$\sim 1-10$ crossing times \cite{nak}. Since the predicted ``frequency
of substructure'' is highly sensitive to the substructure survival
time, the predictions made by the previous studies are too uncertain
to be useful (Kauffmann \& White 1993; Lacey \& Cole 1993).  In our
model, the relevant relaxation timescale is realistically confined to
$\sim 1-2$ crossing times since ${\Delta M /\,\overline{M}}$ is
concerned only with the low order moments in the cluster potential
(i.e. large potential fluctuations containing $\ga$ 20\% of the total
cluster mass).  Moreover, since it is only the expected correlation of
${\Delta M /\,\overline{M}}$ with the low order moments that we
require, and not the actual value, the uncertainty associated with
$\sim 1-2$ crossing times is unimportant for our present study of
$\Omega_0$ (\S \ref{cdmz0}).

The results of our study suggest the following procedure to optimize
constraints on $\Omega_0$ and $\lambda_0$ from analysis of cluster
morphologies quantified in terms of $P_m/P_0$. The present epoch
exhibits the strongest dependence of cluster morphologies on
$\Omega_0$ with $\langle{\Delta
M/\,\overline{M}}\rangle\sim\Omega_0^{1/2}$ -- the higher mass
clusters $M\ga 7\times 10^{14}h^{-1}M_{\sun}$ are slightly preferred
for this purpose. At $z=0$, $\langle{\Delta M/\,\overline{M}}\rangle$
depends only weakly on radius. However, $P_m/P_0$ declines with radius
because the higher order moments give way to the monopole term as the
radius is increased. Hence, any $r\la 1h^{-1}$ would seem best in this
light. (Although not too small to invalidate the connection between
the 2-D and 3-D monopoles -- \S \ref{obs}.)

Although the dependence of $\langle{\Delta M/\,\overline{M}}\rangle$
weakens with increasing redshift, analyzing high-redshift data
separately and combining with $z=0$ data significantly improves the
ability to constrain models. First, simply obtaining independent
cluster samples at medium $(z\sim 0.4)$ and high $(z\sim 0.8)$
redshifts and combining with $z=0$ data increases the dependence on
$\Omega_0$ from $\sim\Omega_0^{1/2}$ at $z=0$ to
$\sim\Omega_0^{1/2}\times\Omega_0^{0.29}\times\Omega_0^{0.23}\sim\Omega_0$.
Second, the evolution of $\langle{\Delta M/\,\overline{M}}\rangle$
with $z$ depends on $\Omega_0$ as well: the ratio of $\langle{\Delta
M/\,\overline{M}}\rangle$ for $z=0.8$ and $z=0$ models goes
approximately as $\Omega_0^{-1/4}$ indicating that the morphologies of
clusters in a low-density universe undergo more rapid evolution than
if $\Omega_0=1$ thereby providing an additional constraint.  Finally,
the radial dependence of $\langle{\Delta M/\,\overline{M}}\rangle$ is
very different for $\Omega_0=0.2,1$ models over radii
$(0.5-1.5)h^{-1}$ Mpc, and we would expect that $P_m/P_0$ would
decrease with radius considerably more rapidly for $z\ga 0.5$ if
$\Omega_0=1$ than if $\Omega_0=0.2$ with or without a cosmological
constant.

Because the dependence of $\Delta M/\,\overline{M}$ on $\Omega_0$
weakens with increasing redshift, X-ray observations of clusters have
a clear advantage over gravitational lensing observations since the
latter are essentially restricted to $z\ga 0.15$. Moreover, the
$P_m/P_0$ computed on the X-ray surface brightness distributions of
clusters have a larger dynamic range than those computed on the
projected mass distributions (\S 4 of Tsai \& Buote 1996), which
probably translates to a greater sensitivity for probing $\Omega_0$
(see end of \S \ref{obs}).  For $z\ga 0.2$, both X-ray and lensing
observations are well suited for probing the evolution of cluster
morphologies with redshift.  However, the theoretical comparison for
lensing data only requires dissipationless simulations which is a
definite advantage over the more computationally expensive
dissipational simulations required for the comparison to X-ray data.

Although we believe at present that further refinements to our model
are not justified given the uncertainty in the relaxation time and the
qualitative nature of the relationship between ${\Delta M
/\,\overline{M}}$ and the low order multipole ratios, there are areas
which could use some attention. A study quantifying the relationship
between ${\Delta M /\,\overline{M}}$,
$\langle(\Phi^{int}_l)^2\rangle^{1/2}/\langle(\Phi^{int}_0)^2\rangle^{1/2}$,
and $P_m/P_0$ using N-body simulations would help clarify predictions
of this model (i.e. determine $c_m$ and $d_m$ -- see \S
\ref{obs}), and will enable a direct comparison to observations of
cluster surface density maps obtained from gravitational lensing and
X-ray data.  If such a study can also significantly limit the range of
relaxation times for low order potential fluctuations, then
investigation of the secondary effects such as non-sphericity and
angular momentum may be warranted.

To illustrate the potential for such studies, let us compare ${\Delta
M /\,\overline{M}}$ to the $P_m/P_0$ computed in \S 4 of Tsai \& Buote
\shortcite{tb} for the dark matter (i.e. $\rho$ as opposed to
$\rho^2$) of clusters formed in a $\Omega_0=1$ CDM simulation. The
even $P_m/P_0$ are nearly proportional to each other such that
$P_2/P_0\approx 100P_4/P_0$, and the mean value of $P_2/P_0$ is $\sim
10^{-4}$.  If we assume the same scaling applies to the monopole,
i.e. $({\Delta M /\,\overline{M}})^2\approx 100P_2/P_0$, then we
estimate ${\Delta M /\,\overline{M}}\approx 0.10$. This value is in
qualitative agreement with the value of $\approx 0.4$ we determined in
\S \ref{cdmz0}. Since the magnitude of ${\Delta M /\,\overline{M}}$
scales approximately linearly with relaxation time, this suggests that
the appropriate relaxation timescale for the low order multipoles is
$<1t_{cross}=0.14/H_0$.

Hence, cluster morphologies would appear to be an attractive means to
probe $\Omega_0$ and $\lambda_0$. Theoretically, cluster morphologies
are related to $\Omega_0$ and $\lambda_0$ in a conceptually
straightforward way, are sensitive to $\lambda_0$ at $z=0$, and are
very insensitive to $\Omega_B$ (\S \ref{cdmz0}) and the cluster bias
factor unlike some other indicators for $\Omega_0$ \cite{dbw}.
Observationally, cluster morphologies are easy to quantify in terms of
$P_m/P_0$ (Buote \& Tsai 1996).  The challenge lies in acquiring
lensing and X-ray data of large samples $(\ga 50)$ of clusters at
various redshifts for comparison to large, high-resolution N-body
simulations of lensing maps and hydrodynamic simulations of X-ray
images.

Alternatively, if N-body simulations confirm that the correlations
between ${\Delta M /\,\overline{M}}$ and the low order $P_m/P_0$ are
tight and mostly independent of $\Omega_0$, then the N-body
simulations will only be needed to initially calibrate this
correlation. In this case subsequent comparison of ${\Delta M
/\,\overline{M}}$ to observations can be achieved rapidly using the
spherical accretion model and allow $\Omega_0$ and $\lambda_0$ to be
determined with considerably less computational expenditure than
required for the standard method.

\section*{Acknowledgments}

We thank P. Natarajan and S. Sigurdsson for comments on the
manuscript. We also thank the anonymous referee for suggesting that we
emphasize the direct application of this model to observations (\S
\ref{obs}).

\appendix

\section{Spherical model for case $\mathbf{\Omega_0} +
\mathbf{\lambda_0} = 1$} 
\label{lambda}

Here we give the formulae for the radius and time at maximum expansion
of spherical shells evolving in a universe where $\Omega_0 + \lambda_0
= 1$ (see \S \ref{sphmod}). Related expressions may be found in
Richstone et al. \shortcite{rlt} for the case of homogeneous density
perturbations, and in Eke et al. \shortcite{ecf}, who adopt the
equations of Peebles \shortcite{p84} which assume a particular initial
condition, $a=r$ at $z\rightarrow \infty$, where $a$ is the expansion
factor and $r$ is the proper radius of the mass shell. We give
expressions explicitly defined for an arbitrary initial epoch.

We begin with the energy integral for a spherical perturbation in a
universe with non-zero cosmological constant $\Lambda \equiv
3H_i^2\lambda_i$,
\begin{equation}
E = \frac{1}{2}\left(\frac{dr}{dt}\right)^2 - \frac{GM(<r)}{r} -
\frac{1}{6}\Lambda r^2,
\end{equation}
which is conserved before shell-crossing. The mass enclosed within
proper radius $r$ is,
\begin{equation}
M(<r) = \frac{4\pi}{3}\rho_b(t_i)(1+\overline{\delta}_i)r_i^3,
\end{equation}
where $\rho_b(t_i)$ is the density of the background universe and
$r_i$ the proper radius at the initial time $t_i$ which has expanded
to radius $r$ at a time $t\le t_m$. The energy evaluated at $t_i$ is,
\begin{equation}
E_i = -K_i\Omega_i\overline{\delta}_i,
\end{equation}
where $K_i=H_i^2r_i^2/2$ is the initial kinetic energy and we have
neglected peculiar velocities and set $\lambda_i=1-\Omega_i$. At
maximum expansion the energy is all potential,
\begin{equation}
E_m = \frac{r_i}{r_m}\left(-K_i\Omega_i\left[
1+\overline{\delta}_i\right] \right) + \left(\frac{r_m}{r_i}\right)^2
\left(-K_i\left[1-\Omega_i\right]\right),
\end{equation}
where $r_m$ is the proper radius at maximum expansion. Since the
energy is conserved we may set $E_m = E_i$ and obtain an equation for
$r_m/r_i$, 
\begin{equation}
\frac{\Omega_i^{-1}-1}{1+\overline{\delta}_i}
\left(\frac{r_m}{r_i}\right)^3 -
\frac{\overline{\delta}_i}{1+\overline{\delta}_i}
\left(\frac{r_m}{r_i}\right) +1 = 0,
\end{equation}
which has the standard cubic solution,
\begin{equation}
\frac{r_m}{r_i} = \frac{2}{\sqrt{3}} \left(
\frac{\overline{\delta}_i}{\Omega_i^{-1}-1} \right)^{1/2}
\cos\left(\frac{\Theta}{3}+\frac{\pi}{3}\right),
\end{equation}
where,
\begin{equation}
\cos \Theta = \frac{3^{3/2}}{2}
\frac{1+\overline{\delta}_i}{\Omega_i^{-1}-1} \left(
\frac{\Omega_i^{-1}-1}{\overline{\delta}_i} \right)^{3/2},
\end{equation}
for $0\le\Theta\le\pi/2$.  Finally, by setting the energy before
maximum expansion equal to the initial energy we obtain an equation
for the time as a function of $r$. The time corresponding to maximum
expansion is thus,
\begin{eqnarray}
\lefteqn{t_m = t_i + 
H_i^{-1}\Omega_i^{-1/2}(1+\overline{\delta}_i)^{-1/2}
\times} \nonumber\\
& &
\int_1^{r_m/r_i} dR\sqrt{R}
\left(\frac{\Omega_i^{-1}-1}{1+\overline{\delta}_i}R^3 -
\frac{\overline{\delta}_i}{1+\overline{\delta}_i}R +1\right)^{-1/2}.
\end{eqnarray}

\end{document}